\documentclass{jpp}
\usepackage{epstopdf, epsfig}

 \usepackage{amssymb}
 \usepackage{bigints}
 \usepackage{amsmath}
 \usepackage{empheq}
 \usepackage{framed}
 \usepackage{color}
 \usepackage{enumerate}
 \usepackage{ulem}
 \usepackage{bm}
 \usepackage{appendix}
 \usepackage{multicol}
 \numberwithin{equation}{section}
 \usepackage{upgreek}
 \usepackage{textgreek}
\usepackage{stmaryrd} 
\usepackage{xfrac}    

\usepackage{caption}
\usepackage{subcaption}

\newcommand{\comment}[1]{}

\newcommand{\be}{\begin{equation}}
\newcommand{\ee}{\end{equation}}
\newcommand{\ba}{\[\begin{aligned}}
\newcommand{\ea}{\end{aligned}\]}
\newcommand{\bea}{\begin{eqnarray}}
\newcommand{\eea}{\end{eqnarray}}
\newcommand{\beann}{\begin{eqnarray*}}
\newcommand{\eeann}{\end{eqnarray*}}
\newcommand{\bs}{\begin{split}}
\newcommand{\es}{\end{split}}

\newcommand*{\cH}{\mathcal{H}}
\newcommand*{\cI}{\mathcal{I}}
\newcommand*{\cJ}{\mathcal{J}}

\newcommand*{\cN}{\mathcal{N}}

\newcommand*{\ep}{\epsilon}

\newcommand*{\B}{\bm{B}}
\newcommand*{\C}{\bm{C}}

\newcommand*{\E}{\bm{E}}
\newcommand*{\F}{\bm{F}}

\newcommand*{\J}{\bm{J}}

\newcommand*{\Q}{\bm{Q}}

\newcommand*{\kv}{\bm{k}}

\newcommand*{\w}{\bm{w}}

\newcommand*{\QDB}{\Q\cdot \dl B}
\newcommand*{\QD}{\Q\cdot \dl }

\newcommand*{\BQ}{\lbr \B\cdot \Q\rbr}

\newcommand*{\BxDpsi}{\B\times \dl \psi}

\newcommand*{\JxB}{\bm{J}\times \bm{B}}

\renewcommand*{\u}{\bm{u}}
\newcommand*{\uB}{\lbr \u\cdot \B\rbr}
\newcommand*{\uQ}{\lbr \u\cdot \Q \rbr}

\newcommand*{\uf}{\bm{u}}

\newcommand*{\jpl}{j_{||}}

\newcommand*{\upr}{\bm{u_\perp}}

\newcommand*{\dl}{\bm{\nabla}}
\newcommand*{\del}{\partial}
\newcommand*{\BD}{\bm{B}\cdot\bm{\nabla}}

\newcommand*{\lbr}{\left(}
\newcommand*{\rbr}{\right)}

\newcommand*{\fb}{\bar{f}}
\newcommand*{\gb}{\bar{g}}
\newcommand*{\vb}{\bar{v}}

\newcommand*{\PhiO}{\Phi^{(0)}}

\newcommand*{\BO}{\B^{(0)}}

\newcommand*{\pOne}{p^{(1)}}

\newcommand*{\PhiOne}{\Phi^{(1)}}

\newcommand*{\BOne}{\B^{(1)}}
\newcommand*{\mBOne}{B^{(1)}}

\newcommand*{\PiOne}{\Pi^{(1)}}
\newcommand*{\uOne}{u^{(1)}}
\newcommand*{\vOne}{v^{(1)}}
\newcommand*{\wOne}{w^{(1)}}

\newcommand*{\rhoOne}{\rho^{(1)}}

\newcommand*\reallywidetilde[1]{\ThisStyle{%
  \setbox0=\hbox{$\SavedStyle#1$}%
  \stackengine{-.1\LMpt}{$\SavedStyle#1$}{%
    \stretchto{\scaleto{\SavedStyle\mkern.2mu\sim}{.5467\wd0}}{.7\ht0}%
  }{O}{c}{F}{T}{S}%
}}

\def\test#1{$%
  \reallywidetilde{#1}\,
  \scriptstyle\reallywidetilde{#1}\,
  \scriptscriptstyle\reallywidetilde{#1}
$\par}
\makeatletter
\newsavebox{\@brx}
\newcommand{\llangle}[1][]{\savebox{\@brx}{\(\m@th{#1\langle}\)}%
  \mathopen{\copy\@brx\mkern2mu\kern-0.9\wd\@brx\usebox{\@brx}}}
\newcommand{\rrangle}[1][]{\savebox{\@brx}{\(\m@th{#1\rangle}\)}%
  \mathclose{\copy\@brx\mkern2mu\kern-0.9\wd\@brx\usebox{\@brx}}}
\makeatother

\shorttitle{3D plasma flows} 

\title{Steady plasma flows in a periodic non-symmetric domain}

\shortauthor{H.Weitzner,W. Sengupta,}
\author{Harold Weitzner \aff{1}, Wrick Sengupta \corresp{\email{wricksg@gmail.com}} \aff{2,3}}

\affiliation{\aff{1}Courant Institute of Mathematical Sciences, New York University, New York, New York 10012, USA
\aff{2}Department of Astrophysical Sciences, Princeton University, Princeton, NJ, 08543
\aff{3}Princeton Plasma Physics Laboratory, Princeton, NJ, 08540
}

\begin{document}

\maketitle

\begin{abstract}
Steady plasma flows have been studied almost exclusively in systems with continuous symmetry or in open domains. In the absence of continuous symmetry, the lack of a conserved quantity makes the study of flows intrinsically challenging. In a toroidal domain, the requirement of double-periodicity for physical quantities adds to the complications. In particular, the magnetohydrodynamics (MHD) model of plasma steady-state with the flow in a non-symmetric toroidal domain allows the development of singularities when the rotational transform of the magnetic field is rational, much like the equilibrium MHD model. In this work, we show that steady flows can still be maintained provided the rotational transform is close to rational and the magnetic shear is weak. We extend the techniques developed in carrying out perturbation methods to all orders for static MHD equilibrium by Weitzner (Physics of Plasmas 21, 022515 (2014)) to MHD equilibrium with flows. We construct perturbative MHD equilibrium in a doubly-periodic domain with nearly parallel flows by systematically eliminating magnetic resonances order by order. We then utilize an additional symmetry of the flow problem, first discussed by E. Hameiri in (J. Math. Phys. \textbf{22}, 2080 (1981) Sec. III), to obtain a generalized Grad-Shafranov equation for a class of non-symmetric three-dimensional MHD equilibrium with flows both parallel and perpendicular to the magnetic field. For this class of flows, we can obtain non-symmetric generalizations of integrals of motion, such as Bernoulli's function and angular momentum. Finally, we obtain the generalized Hamada conditions, which are necessary to suppress singular currents in such a system when the magnetic field lines are closed. We do not attempt to address the question of neoclassical damping of flows.
\end{abstract}

\section{Introduction \label{sec:intro}}
Flows in magnetic confinement devices are known to have a  strong influence in suppressing turbulence and affecting neoclassical \citep{Hinton_Wong_1985_neoclassical_transport_rotating,helander1998neoclassical} and turbulent transport \citep{burrell1997_ExB_effects}, as well as stability \citep{chu_Greene_1995effect_toroidal_flow}. In particular, sheared flows like zonal flows can reduce turbulent transport by shearing the turbulent eddies \citep{lin_Hahm_1998_Zonal_flow_suppress_turbulent,fujisawa2008_zonal_flow_review,terry2000_turb_suppress_shear_flow}. Study of the interaction of plasma flows and turbulence is critical \citep{stroth2011_plasma_flow_interaction,groebner_Burrel_1990role} in understanding the edge region of tokamaks and the L-H transition. It has been experimentally verified that the radial electric field plays a key role in the transition \citep{burrell1999tests_ExB_turbulence}.
Flows can also have applications in healing magnetic islands in stellarators \citep{hegna2012_flow_healing_island}. Flows are also an essential object of study in space and astrophysical plasmas \citep{Goedbloed2004_MHD_book,beskin2009mhd,Istomin1996stability,Davis2020magnetohydrodynamics_AGN}. 

If the  flow speed is assumed on the order of the sound speed, ideal MHD is a good description of plasmas since gyroradius corrections do not enter \citep{morozov1980steady_flow,freidberg2014ideal,Goedbloed2004_MHD_book}. The theory of steady axisymmetric plasma flows is well developed \citep{hameiri1983_MHD_flow_equilibrium,hassam1996_NF_poloidal_sonic,guazzotto2005_MHD_toroidal_poloidal_flow,mcclements2010steady,tasso1998axisymmetric,Abel_2013_multiscale_GK,beskin1997axisymmetric}. Toroidal angular momentum conservation, which follows from Noether's theorem, facilitates the reduction of axisymmetric flows to the study of a modified Grad-Shafranov (GS) equation which includes the effect of steady plasma flow \citep{hameiri1983_MHD_flow_equilibrium,hassam1996_NF_poloidal_sonic,tasso1998axisymmetric,Goedbloed1997stationary_symmetric_MHD_flows,beskin1997axisymmetric}. Similarly, for plasma systems with helical symmetry, a helical Grad-Shafranov with large flows can be obtained \citep{andreussi_Morrison_2012hamiltonian,Villata_Tsinganos_1993exact_helical_MHD_eqb}. Global solutions of such flow modified Grad-Shafranov equations
provide useful models of plasma jets in space \citep{bogoyavlenskij2000_Exact_astro_jets_MHD_Eqb,Villata_Ferrari_1994exact_helical_astro_jets, bogoyavlenskij2000_helical_asttro_jets,beskin1997axisymmetric}.

While the standard GS equation for static equilibrium is always elliptic, the flow-modified GS equation can be elliptic, parabolic, or hyperbolic, depending on the flow's Mach number. Therefore, even with symmetry, complicated flows such as transonic flows \citep{morawetz1985weak} that modify the characteristics in a given domain can appear \citep{lifschitz_Goedbled1997transonic}.

There is considerable interest in understanding the effects of symmetry breaking three-dimensional perturbations and stability of three-dimensional MHD flows in astrophysical plasmas \citep{Igumenshchev2003three,Mckinney2009stability,Istomin1996stability,Igumenshchev2008_MAD_numerics}. In particular, 2D and 3D simulations of magnetically arrested disks show very different behaviors \citep{White2019_MAD_resolution,Igumenshchev2008_MAD_numerics}. However, high-resolution 3D simulations are required to carefully capture the effects of symmetry-breaking \citep{White2019_MAD_resolution}. 

In the absence of symmetry, reduction to a Grad-Shafranov like equation fails in general. Non-symmetric toroidal MHD equilibrium with flows share the same mathematical difficulty as the static non-symmetric MHD equilibrium, namely, the appearance of singularities on rational surfaces. 

Although large steady shear-flows are desirable in various confinement geometries, steady flows show a strong alignment with a symmetry direction when present \citep{helander2007rapid,helander_Simakov_PRL_2008_intrinsic_ambipolarity,Helander2014,spong2005generation}. In axisymmetric tokamaks, the persistent flow is in the toroidal symmetry direction \citep{Throumoulopoulos_tasso_Weitzner2006_nonexistence_purely_poloidal_flow,tasso1998axisymmetric}. While the toroidal angular momentum in a tokamak is typically conserved over several inverse collision frequency \citep{Helander2012stellarator_v_tokamak}, poloidal flows are damped due to neoclassical effects.  Axisymmetry makes tokamaks intrinsically ambipolar, i.e., the radial fluxes of ions and electrons are equal and independent of the radial electric field in the leading order gyroradius expansion, leading to flow in the symmetry direction being unconstrained. 

In the absence of intrinsic amibipolarity, the rotation is damped to the value required for ambipolar radial transport and typically leads to flows of the order of diamagnetic flows \citep{helander_Simakov_PRL_2008_intrinsic_ambipolarity, Helander2014}.
Plasma rotation in tokamaks and stellarators, therefore, differ significantly because stellarators do not, in general, have a symmetry direction and are not intrinsically ambipolar \citep{Helander2012stellarator_v_tokamak}. The only class of stellarators that have been shown to support large flows are quasisymmetric \citep{tessarotto1996,canik_HSX_2007_improved_neoclassical,spong2005generation,helander2007rapid,helander_Simakov_PRL_2008_intrinsic_ambipolarity}. In a quasisymmetric stellarator, continuous symmetry comes from the symmetry of the magnitude of the magnetic field. As a result of the symmetry, such stellarators are also approximately intrinsically ambipolar \citep{helander_Simakov_PRL_2008_intrinsic_ambipolarity}. However, it is difficult, if not impossible, to obtain exact global quasisymmetry in a volume \citep{Garren1991a,Landreman_Sengupta_ho_2019}. The breakdown of quasisymmetry leads to flow damping \citep{spong2005generation,Simakov_Helander_2011plasma_rotation_QS,helander_Simakov_PRL_2008_intrinsic_ambipolarity}. In the literature, it has been further argued in the literature \citep{sugama2011_QS_large_flow,Simakov_Helander_2011plasma_rotation_QS,tessarotto1996} that quasisymmetry is not enough to accommodate the large centrifugal forces from sonic flows, and one needs axial/helical symmetries. 

As a result of the difficulties imposed by the three-dimensional geometry, a lack of symmetry that leads to magnetic resonances on rational surfaces and stringent restrictions on plasma flows from neoclassical physics, a complete mathematical description of the characteristics of a generic 3D ideal MHD equilibrium, let alone stability of such systems is not available in the standard literature.
The analytic description of nonsymmetric 3D flows has employed various  approximations such as incompressibility \citep{kamchatnov1982topological_soliton_mhd} or large aspect ratio \citep{kovrizhnykh1989_3D_avg_MHD_flow,kovrizhnykh1980_3D_avg_MHD} have been invoked. 

Several variational formalisms for 3D plasma flow problem exist \citep{hameiri1998variational,greene1969variational,Ilgisonis_Pastukhov_2000_variational,Vladimirov_Moffatt_1995_variational_MHD}. A nontrivial result from the variational formalism is the proof that under certain conditions ideal MHD allows more than one kind of cross-helicity \citep{hameiri1981spectral_C,Ilgisonis_Pastukhov_2000_variational,Vladimirov_Moffatt_1995_variational_MHD}. We shall call the additional symmetry vector associated with the conserved quantity Hameiri's vector because it was first discovered in \citep{hameiri1981spectral_C} and later derived independently in \citep{Ilgisonis_Pastukhov_2000_variational} and \citep{Vladimirov_Moffatt_1995_variational_MHD} (see discussion in \citep{hameiri1998variational} and \citep{Ilgisonis_Pastukhov_2000_variational}). A major disadvantage of the flow-modified variational formalism is that, unlike the static limit, the energy functional for the flow problem does not possess a minimum unless the equilibrium flow is small enough or is parallel to the magnetic field \citep{hameiri1998variational}.

In this work, instead of addressing the flow damping problem in a 3D system, we take a step back and analyze the ideal MHD equilibrium with steady flows in a non-symmetric geometry. Our goal here is to construct a class of perturbative solutions to non-symmetric ideal MHD with steady flows and obtain consistency conditions for steady flows when the magnetic fields close on themselves. The flows that we consider are larger than diamagnetic flows but are not sonic. We show that the same procedure that allowed us to obtain non-symmetric ideal MHD static equilibrium in a low shear stellarator can be generalized to MHD with such flows. In particular, we highlight the crucial role of closed field lines \cite{Grad1971plasma,weitzner2014ideal, Weitzner2016} in avoiding magnetic resonances and going to arbitrarily high order in the perturbative analysis. We also show that for a special class of flows that possess the additional symmetry, the Hameiri's vector, one can obtain a generalized Grad Shafranov equation (GGS) similar to the quasisymmetric generalized Grad Shafranov equation first derived in \citep{Burby2020} (see also related works by \cite{burby2020_GGS,Constantin2020}). The GGS reduces to the standard flow-modified Grad-Shafranov equation in symmetric geometries. We explicitly obtain the necessary conditions that need to be satisfied on rational surfaces. If these conditions are not satisfied current singularities can develop on rational surfaces \citep{Soloviev_Shafranov_v5_1970plasma, Grad1967toroidal_confinement,Boozer_coords_1981}.

The outline of the paper is as follows. First, in section \ref{prelim}, we study the ideal MHD equilibrium with flows and identify the characteristic surfaces. Then, in section \ref{Linearization} we study the perturbation of a class of exact solutions and obtain the dispersion relation. Then, we outline in section \ref{nearly_parallel} how the perturbation theory can be carried out to all orders for nearly parallel flows. Next, we utilize Hameiri's vector to derive the GGS and the necessary conditions on rational surfaces in section \ref{sec:GGS_C}. Finally, we discuss the implications of our results in section \ref{sec:discussion}.
 
\section{Preliminaries}
\label{prelim}
We employ the standard model of ideal magnetohtdrodynamics to represent the plasma in steady flow state but we add the simplifying assumption that the entropy is a constant in the entire plasma. The system is
\begin{subequations}
\begin{align}
    \dl\cdot (\rho \uf )&=0 \label{density_eqn}\\
    \rho \uf \cdot \dl \uf +\dl p &=\J \times \B,\quad \J=\dl\times \B \label{momentum_eqn}\\
    \dl \cdot \B&=0 \label{divB}\\
    -\E=\dl \Phi &= \uf \times \B. \label{Efield}
\end{align}
\label{ideal_MHD_flows}
\end{subequations}
It is often convenient to replace \eqref{Efield} by the equivalent expression
\begin{align}
    \dl \times \lbr \uf \times \B\rbr=0 \label{NoEfield} \tag{\ref{Efield}*}
\end{align}
While \eqref{Efield} and \eqref{NoEfield} are equivalent, the form \eqref{Efield} makes explicit the need to select specific EMFs, equivalently periods associated with the potential $\Phi$. As noted we add the assumption
\begin{align}
    p=p(\rho) \label{pressure_eqn}
\end{align}
to close the system. For concreteness, we shall assume an adiabatic equation of state of the form $p\sim \rho^\gamma$, where $\gamma$ is the ratio of specific heats. If we introduce
\begin{align}
    \Pi\equiv p+\frac{1}{2}\B\cdot\B \label{Pi_eqn}
\end{align}
then an equivalent form for \eqref{momentum_eqn} is 
\begin{align}
    \rho \uf \cdot \dl \uf +\dl \Pi =\BD \B \label{Pi_momentum_eqn} \tag{\ref{momentum_eqn}*}
\end{align}
 We start with a formal mathematical exploration of the system
\eqref{density_eqn}-\eqref{pressure_eqn}. It is clear there are eight scalar first-order differential equations for the eight scalar unknowns $\rho,\Phi,\uf,\B$. If we were to use \eqref{NoEfield} instead of \eqref{Efield}, there would be one fewer unknown but also a hidden solvability condition for \eqref{NoEfield} that adds an unknown and leads back to \eqref{Efield}. 
We now wish to determine the ``type" of the system, elliptic, hyperbolic, parabolic, mixed, or whatever may be. 
A closely related issue is the description of the standing waves in the linearized system. We linearize about a constant state for $\rho, \u$ and $\B$  and look for waves propagating as $\exp{(i \bm{k} \cdot \bm{x})}$. Such solutions are special cases of simple waves in fluid dynamics \citep{Courant_Friedrichs_1999supersonic}. With $c^2\equiv dp/d\rho$, we obtain the following linear system 

\begin{subequations}
\begin{align}
    (\kv \cdot \uf) \delta \rho + \rho (\kv \cdot \delta\uf) &=0 \label{17a}\\
    \rho (\kv \cdot \uf)\delta \uf +c^2 \kv \delta \rho&= \delta \B (\kv \cdot \B)-\kv (\delta \B \cdot \B) \label{17b}\\
    \kv \times (\delta \uf \times \B)+\kv \times (\uf \times \delta \B) &=0 \label{17c}\\
    \kv \cdot \delta \B &=0. \label{17d}
\end{align}
\label{k_pertb_eqn_system}
\end{subequations}
It follows easily that given the  first order perturbations $\del \rho$, $\delta \u$ and $\delta \B$ , they satisfy the system (2.7a)-(2.7d)

From \eqref{17c} we find 
\begin{align}
\delta \uf (\kv \cdot \B)-\B (\kv \cdot \delta \uf)- \delta \B (\kv \cdot \uf) =0
    \label{17cs} \tag{\ref{17c}*}
\end{align}
while \eqref{17b} together with \eqref{17a} yields
\begin{align}
    \rho (\uf \cdot \kv)^2 \delta \uf -\rho \kv c^2 (\kv \cdot \delta \uf)= \lbr \kv \times \lbr \delta \u (\kv \cdot \B)-\B (\kv \cdot \delta \u) \rbr \rbr \times \B
    \label{pre_disp_eqn}
\end{align}
Taking components of \eqref{pre_disp_eqn} in $\bm{k}$ and $\B$ directions, we obtain
\begin{subequations}
\begin{align}
    \rho \lbr (\kv \cdot \uf)^2-c^2k^2\rbr \kv \cdot \delta \uf = k^2 B^2 (\kv \cdot \delta \uf)-k^2 (\kv \cdot \B)(\B\cdot \delta \uf)\\
    \rho  (\kv \cdot \uf)^2(\B\cdot \delta \uf)-\rho (\kv\cdot \B)c^2 (\kv \cdot \delta\uf) =0.
\end{align}
\end{subequations}
A simple calculation shows that the condition for a solution is the “dispersion relation” : 
\begin{align}
    \lbr \rho  (\kv \cdot \uf)^2\rbr^2 -\rho  (\kv \cdot \uf)^2\lbr \rho c^2 +B^2\rbr k^2 +k^2 (\kv \cdot \B)^2\rho c^2=0
    \label{The_disp}
\end{align}

We observe that this condition is satisfied identically for all vectors $\kv$ perpendicular to $\u$ and $\B$. The expression also shows connections to the fast-slow wave combination found in ideal magnetohydrodynamic wave propagation studies.  Note, however, that no Alfv\'en wave appears.  Moreover, the magnitude of $\kv$ does not appear in the system; only its direction occurs. We study the dispersion relation \eqref{The_disp} and its solutions in much more detail in Appendix \ref{app:The_disp}. In particular, we show that real physical solutions of \eqref{The_disp} exist.
  
An alternate interpretation of the relation is also significant.  We might pose the question: What are the characteristic surfaces for the steady flow system? Characteristic surfaces are defined by the following property.  Suppose one specifies $\rho, \u$ and $\B$ on a given surface.  Can one then determine the normal derivative of these quantities on that surface?  If so, the surface is not characteristic.  On a characteristic surface, the flow variables are not arbitrary; they must be related. These surfaces define important properties of any solution. 
Suppose $\chi=0$ is such a surface and define  $\kv$ as the gradient of $\chi$. Hence  $\kv$ is normal to the surface.  If one further defines $\delta \rho, \delta \u$ and $\delta \B$ as the gradients of these variables dotted into the normal to the surface, then these quantities satisfy the system (2.7), except that there would be inhomogeneous terms added to the equations, terms given solely in terms of the variables on the given surface.  With this interpretation of the system, the relation is the condition that one cannot give the unknown functions arbitrarily on the surface in question.  Such peculiar surfaces, the characteristic surface, are thus defined by \eqref{The_disp}. 

A physical problem without such real surfaces is typically elliptic.  When they exist, there are usually hyperbolic properties of the system as well. Their appearance indicates the complexity of steady magnetohydrodynamic flow. Flows with both elliptic and hyperbolic characteristics can occur in ideal magnetohydrodynamic equilibria. The mixed nature (neither fully elliptic nor hyperbolic) of MHD leads to difficult unresolved mathematical issues.

\section{A simple linearized problem}
\label{Linearization}
In order to gain some insight into the nature of steady flow states we examine a simple linearized problem. We start from an exact solution of the system \eqref{density_eqn}-\eqref{Efield}
\begin{subequations}
\begin{align}
    \B &= (0,f(x),g(x))\\
    \uf &= (0,v(x),w(x))\\
    p&=p(x)\\
    \Pi&=p(x)+\frac{1}{2}(f^2+g^2)\\
    -\E&=\Phi_{,x}= v(x)g(x)-w(x)f(x).
\end{align}
\label{eqb_exact_soln}
\end{subequations}
We linearize about the state and introduce the perturbed variables with the structure 
\begin{subequations}
\begin{align}
    \BOne&= (\mBOne_x(x) \sin{(my + nz)},\mBOne_y(x) \cos{(my + nz) },\mBOne_z(x) \cos{(my + nz) })\\
     \bm{\u}^{(1)}&= (\uOne_x(x) \sin{(my + nz)},\uOne_y(x) \cos{(my + nz) },\uOne_z(x) \cos{(my + nz) })\\
     (\pOne,\PiOne,\PhiOne)&= (\pOne(x),\PiOne(x),\PhiOne(x))\cos{(my+nz)}
\end{align}
\label{linear_amplitudes}
\end{subequations}
We next give the linearized system of equations for the many amplitudes in \eqref{linear_amplitudes}. The two divergent relations are

\begin{subequations}
\begin{align}
    \mBOne_{x,x}-(m \mBOne_y + n \mBOne_z)=0 \label{23a}\\
   - \rhoOne( m v(x)+n w(x))+(\rho(x)\uOne(x))_{,x}-\rho(m \vOne+n \wOne) =0\label{23b}
\end{align}
\label{divergent_reln}
\end{subequations}
while $\PiOne$ is specified as
\begin{align}
    \PiOne= c^2 \rhoOne+f(x) \mBOne_y +g(x) \mBOne_z
    \label{24}
\end{align}
where $c^2$ is the sound speed $dp/d\rho$ at the corresponding value of $\rho(x)$. Ohm's law reads
\begin{subequations}
\begin{align}
    \PhiOne_{,x}&=\vOne g + v \mBOne_z - \wOne f - w \mBOne_y \label{25a}\\
    -m \PhiOne&= w(x) \mBOne_x - \uOne g(x) \label{25b}\\
    -n \PhiOne&= -v(x) \mBOne_x + \uOne f(x) \label{25c}
\end{align}
\label{PhiOne_eqn}
\end{subequations}

Finally the three momentum balance equations are
\begin{subequations}
\begin{align}
   \uOne(m v+n w)\rho(x)+\PiOne_{,x}&=\mBOne_{x}(m f + n g) \label{26a}\\
  \rho \lbr \uOne v'-(m v+n w)\vOne\rbr -m \PiOne&= \mBOne_{x}f'(x)-(m f+n g)\mBOne_y \label{26b}\\
  \rho \lbr \uOne w'-(m v+n w)\wOne\rbr -n \PiOne&= \mBOne_{x}g'(x)-(m f+n g)\mBOne_z \label{26c}
\end{align}
\label{momentum_balance_eqn}
\end{subequations}
We introduce the notation 
\begin{align}
\fb= m f + n g, \quad \vb = m v + n w
    \label{notation}
\end{align}
We see that $\rhoOne$ occurs only in \eqref{23b} and \eqref{24}. It is convenient to eliminate $\rhoOne$ from the problem and we replace (24) by 
\begin{align}
    \vb \PiOne= c^2\lbr p(m \vOne + n \wOne)-(p \uOne)_{,x}\rbr +f(x) \mBOne_y +g(x)\mBOne_z
    \label{24s}\tag{\ref{24}*}
\end{align}
Provided the series retains the structure in the angles given in (\ref{linear_amplitudes}) in all orders $m=n=0$ is never possible in the two equations: Ohm's law (\ref{25b}) and (\ref{25c}). Thus, we may replace \eqref{25b},\eqref{25c} by the following
\begin{subequations}
\begin{align}
    -\vb \PhiOne&= \uOne\lbr v g - w f\rbr = \uOne \PhiO_{,x} \label{25bs}\tag{\ref{25b}*}\\
    -\fb \PhiOne&= \lbr u f - v g\rbr \mBOne_x =-\mBOne_x   \PhiO_{,x}
    \label{25cs}\tag{\ref{25c}*}
\end{align}
\label{VbfbPhiOne}
\end{subequations}
We may carry out similar reductions of pressure balance and from \eqref{25b} and \eqref{25c} we obtain 
\begin{subequations}
\begin{align}
    \rho \uOne \vb'-\rho \vb (m \vOne + n \wOne)-(m^2+n^2)\Pi= \mBOne_x \fb' -\fb (n \mBOne_y  - m \mBOne_z) \label{26bs} \tag{\ref{26b}*}\\
    \rho \uOne (n v' -m w')-\rho \vb (n \vOne -m \wOne)= \mBOne_x (n f - m g ) -\fb (n \mBOne_y  - m \mBOne_z) \label{26cs} \tag{\ref{26c}*}
\end{align}
\label{rhouOne_vprime}
\end{subequations}
The $x$ component of \eqref{26a} has the simpler form 
\begin{align}
(v g - w f)\PiOne_{,x} = (\fb^2+\gb^2)\PhiOne
    \label{26as}\tag{\ref{26a}*}
\end{align}
At this point we have five equations (\ref{23a},\ref{25a},\ref{24s}) and (\ref{26bs},\ref{26cs}) in which we may eliminate $\uOne$ and $\BOne_x$ by means of (\ref{25bs},\ref{25cs}) and with the unknowns $(\PhiOne, \PhiOne_{,x},\vOne,\wOne,\mBOne_y,\mBOne_z)$ and $\PiOne$. Hence if we consider $\PhiOne$ and $\PiOne$ as given then we may solve for $\PhiOne_{,x}$. Thus, we obtain a first order homogeneous linear differential equation for $\PhiOne$ of the form $$\PhiOne_{,x}= a(x) \PhiOne +b(x) \PiOne.$$ However, the construction fails with the vanishing of the determinant of the system of five equations for $(\PhiOne_{,x},\vOne,\wOne,\mBOne_y,\mBOne_z)$ in terms of $\PhiOne$ and $\PiOne$. The zeros of this determinant constitute the resonant singularities of the system. We then obtain the determinant of this system by replacing $\mBOne_{,x}$ and $\uOne$ by $\PhiOne$ and ignoring the terms proportional to $\PhiOne$ or $\PiOne$. We find
\begin{align}
\Delta =
    \begin{vmatrix}
    \frac{\bar{f}}{\PhiO_{,x}} \quad \quad 0 \quad\quad\quad 0 \quad \quad-m \quad\quad-n\\
    1\quad\quad -g \quad\quad\quad f \quad\quad\quad w \quad\quad -v\\
    0 \quad\quad\quad \rho \bar{v}\quad\quad\quad 0\quad\quad -\bar{f}\quad\quad\quad 0\\
    0\quad\quad \quad 0\quad\quad\quad \rho \bar{v}\quad\quad\quad 0 \quad\quad -\bar{f}\\
    -\frac{\rho c^2 \bar{v}}{\PhiO_{,x}}\quad \rho c^2 m \quad\:\: \rho c^2 n\quad\quad f\bar{v} \quad\quad g\bar{v}
    \end{vmatrix}
    \label{Det}
\end{align}
and after straightforward reduction we obtain 
\begin{align}
    \Delta= (\rho \vb^2-\fb^2)(\rho \vb^2(c^2+(f^2+g^2)/\rho)-c^2\fb^2)
    \label{28}
\end{align}
For the particular background state, resonances are associated with Alfv\'en waves, the first factor, and a particular form of the fast-slow wave, the second factor.

It is clear that for a general equilibrium there will be resonances when $\vb=\pm \fb/\sqrt{\rho}$ or $\pm \fb c/\sqrt{c^2\rho + (f^2+g^2)}$. For wide ranges of the ratio of the mode indices, $m/n$ resonant surfaces will appear. Just as it was possible to find formal expansions in mode amplitudes of ideal MHD equilibrium, it may be of interest to explore the possibility of formal expansions of flows with no resonant singularities.  

In a more mathematically precise language we can state the main idea as follows. One starts with a homogeneous linear system of five equations in seven unknowns of the form $A X = 0$, where $X$ is a 7-component vector and $A$ is a 5x7 rectangular matrix. Next one write $X = (x,y)$, where $x$ denotes the first 5 components of X, and $y$ denotes the last two components, $\Phi^{(1)}$ and $\Pi^{(1)}$. Then the value of $y$ is fixed, which leads to the inhomogeneous linear system $ax = b$, where $a$ is the restriction of $A$ to the subspace $y=0$ and $b = A(0,y)$. When $\text{det}(a)$ is non-zero, one therefore gets expressions for each component of $x$ that are linear in $y$. When  $\text{det}(a)=0$, there is no solution for $x$ in general unless $b$ is in the image of $a$. The quantity $\Delta$ in eq. \eqref{Det} is just that $\text{det}(a)$.

Such a possibility depends on arranging the state so that it is in resonance everywhere or nowhere. Note that when $\Delta = \text{det}(a)=0$, there is still a solution for $x$ provided $y$ can be adjusted so that $A(0,y)$ is in the image of $a$. Such a situation is possible for parallel flows, i.e., flows where $\B$ and $\u$ are parallel, or 
\begin{align}
    \u= (0,v(x),w(x))=\lambda(x)(0,B_y(x),B_z(x)).
    \label{29}
\end{align}
In this case $\vb=\lambda \fb$, so that \eqref{28} becomes 
\begin{align}
    \Delta= \fb^2\lbr \rho (\lambda(x))^2-1\rbr \lbr \rho (\lambda(x))^2(c^2+(f^2+g^2)/\rho)-c^2 \rbr =0
    \label{30}
\end{align}
If $\lambda(x)$ is chosen so that the second or the third term vanishes, then $\Delta$ is identically zero for every mode. Such flows may well be of some interest, but they seem extremely pathological and do not appear to be realizable. However, the case $\fb=0$ is of more interest and similar to the equilibrium case. In particular we assume the second factor never vanishes and $\fb=0$ for $m=M, n=N$ where $M$ and $N$ are relatively prime. Thus, we assume
\begin{align}
    \BO=(0,N,-M)\mu(x) \label{31}
\end{align}
where $\mu(x)$ is an arbitrary function, while $\u$ is given by (\ref{29},\ref{31}). We expand in amplitude about this state and describe expansions to all order in the amplitude. 

\section{Nearly parallel flows}
\label{nearly_parallel}
In the last section we showed that perturbations of steady flows are typically subject to appearances of singularitiies on particular surfaces. However, for parallel flows given by (\ref{29},\ref{31}) singular surfaces need not appear. In this section we examine the flows with these properties and we lay out the argument that one may construct a formal series solution to all orders without the appearance of any singularities when the lowest order system is a parallel flow. We do not give the full details of the proof of expansions to all orders but we consider the conclusion valid. The process closely parallels the development in \citep{weitzner2014ideal}, where such a formalism is fully developed.

The expansion parameter of the series is the amplitude of the flow and field components. We assume that the lowest order field and flow state is given by (\ref{29},\ref{31}). In first order we add fields and flows which is a sum of terms given by \eqref{linear_amplitudes} subject to the condition that none of them is resonant. We may then continue to construct an expansion order by order. The series we construct are assumed to have the same structure of angular dependence in all orders. The linearized system used to determine the characteristics of the system, closely parallels \eqref{k_pertb_eqn_system}.

In this expansion at each order we must solve a system similar to \eqref{23a},\eqref{24s},\eqref{25a},
\eqref{25bs},\eqref{25cs},\eqref{26a},\eqref{26bs},\eqref{26cs} where sums of products of lower order terms may appear added to the right hand side of the equations. When the mode is not resonant so that $\Delta$ given by \eqref{30} is not zero, one may solve the linear homogeneous system. Our task is to indicate how one should be able to select series so that order by order there is no net resonant term arising in the combination of sums of products of lower orders which form the inhomogeneous terms of the generalization of the system.

Before we address the central issue in this work we must characterize the structure of the resonance more fully. In resonance the quantities $\vb$ and $\fb$ are identically zero. Thus, from \eqref{25bs},\eqref{25cs}, \eqref{26as} and \eqref{VbfbPhiOne}
we find that 
\begin{subequations}
\begin{align}
    u^{1}(x)&=B_x^{(1)}=0
    \label{u1Bx1}\\
    \Phi^{(1)}&=0 \label{PhiOneis0} \quad \quad (\text{from \eqref{VbfbPhiOne}})\\
    \Pi^{1}&=0 \label{PiOneis0} \quad \quad (\text{from \eqref{26as}}).
\end{align}
\label{resonant_eqn1}
\end{subequations}
Finally, the system is closed with the two conditions
\begin{subequations}
\begin{align}
    m B^{(1)}_y + n B^{(1)}_z &=0\\
    m v^{(1)}+ n w^{(1)}&=0
\end{align}
 \label{resonant_eqn2}
\end{subequations}
The state given by \eqref{resonant_eqn1}, \eqref{resonant_eqn2} is the resonant mode in any order, provided
\begin{align}
    \frac{m}{n}=\frac{M}{N}.
    \label{resonant_eqn3}
\end{align}

We finally turn to the inhomogeneous system where we expand the solution to some order in the amplitude say order $P$. We obtain a system of the form \eqref{23a},\eqref{24s}, \eqref{PhiOne_eqn},\eqref{momentum_balance_eqn} where the index one is replaced by $P$ and inhomogeneous terms are added to the right hand sides of the equations. These terms are sums of products of terms of order lower than $P$. We can solve the system provided that there is no net contribution at a resonant term, for if there were such a term we could not guarantee the construction of a periodic solution of the system. We must show that the series can be arranged so that no net inhomogeneous term is present.

We modify the structure of the solution in orders $(P-2)$ and $(P-1)$ so that no net singular terms arise in order $P$. We assume that before modification there are terms in order $P$ with $m$ and $n$ satisfying \eqref{resonant_eqn3}. We add a resonant term in order $(P-2)$ with angular dependence $\exp{i(\mu y +\nu z)}$ and undetermined $x$ dependence. There are two such independent additions $B_y^{(P-2)}$ and $v^{(P-2)}$, say. We assume that there are non-resonant terms in the first order of the form \eqref{linear_amplitudes}. When these terms beat against each other there will be terms in order $P-1$ with angular dependence $\exp{i((\mu\pm m) y +(\nu\pm n) z)}$. Finally, in order $P$, terms from order $(P-1)$ beat again against the first order terms with angular structure $\exp{i(m y +n z)}$. We choose the as yet undetermined $x$ dependence so as to satisfy the two constraints. The quantities $u^{(P)},B_x^{(P)},\Phi^{(P)},\text{ and }\Pi^{(P)}$ which were all zero in the solution of the homogeneous data problem are all non-zero and are determined.

\section{Analysis of steady flows that admit an additional symmetry vector }\label{sec:GGS_C}
We have so far outlined how one can perturbatively construct non-symmetric steady plasma flows in ideal MHD. We considered flows that are nearly parallel to avoid resonances. Compared to non-symmetric MHD equilibrium without flows, the steady flow system \eqref{ideal_MHD_flows} is far more complicated due to extra equations and nonlinearities from the flow variables. In particular, it is not clear if the resonances can be avoided. Fortunately, analytical progress can be made if the steady flows possess a symmetry vector, the Hameiri vector \citep{hameiri1983_MHD_flow_equilibrium}, which is closely related to the quasisymmetry vector \citep{Burby2020,Rodriguez2020a}. In the following, we employ a more formal approach to investigate non-symmetric 3D flows that come equipped with the Hameiri vector. We shall now state our goals and the main results obtained in this section.

Our primary objective is to investigate the properties of a class of non-symmetric 3D flow system that has flow components both parallel and perpendicular to the magnetic field and admits  Hameiri's symmetry vector. Under an additional assumption that the density does not change along this vector, we show that non-symmetric generalizations of Bernoulli's law, angular momentum conservation, a generalized Grad-Shafranov (GGS) equation, and associated generalized Hamada conditions can be obtained. Therefore, our main results show that the description of these special non-symmetric flows parallels symmetric flows.

The organization of this section is as follows. We begin with a discussion of the Hameiri vector and point out its close connections with the quasisymmetry vector in section \ref{subsec:Hameiri_C_and QS}. We then discuss the key assumptions that we make in order to carry out the analysis in \ref{subsec:basic_assumptions}. We derive the generalizations of Bernoulli's law in section \ref{subsec:gen_Bernoulli}.  We show that a generalized Grad-Shafranov equation can be constructed following the approach of \cite{Burby2020} in section \ref{subsec:GGS_flow}. We discuss the generalization of Hamada conditions in section \ref{subsec:last_Hamada} and summarize the results in section \ref{subsec:summary}.

\subsection{ Hameiri's vector $\textbf{C}$ and weak quasisymmetry }\label{subsec:Hameiri_C_and QS}

We briefly summarize a key result due to Hameiri, which is essential to our work.  Assuming a steady flow with nested toroidal flux surfaces labelled by $\psi$, \cite{hameiri1983_MHD_flow_equilibrium} showed that ideal MHD allows an additional cross-helicity of the form $\int d^3\bm{r}\:\u\cdot\C$, where $d^3\bm{r}$ denotes volume integral over 3D space, if there exists a vector $\C$ such that
\begin{align}
    \C\cdot \dl \psi=0,\quad  \dl \cdot \C =0, \quad \dl \times \lbr\B \times  \frac{1}{\rho}\C  \rbr=0, \quad \dl \times \lbr \u \times \C \rbr=0.
    \label{C_conditions}
\end{align}
Hameiri has shown that in axisymmetry, $\C$ is $R^2\dl \varphi$ in the usual $(R,Z,\varphi)$ cylindrical coordinates used in tokamak literature. Furthermore, from \eqref{C_conditions} we can see that the vector $\C$ is analogous to a magnetic field since it is frozen-in with the flow and lies on flux surfaces.

We shall make an additional assumption that the density is constant in the direction of the Hameiri vector i.e. $\C\cdot\dl \rho=0$. Writing $\C=\rho \Q$, we get an equivalent set of conditions on $\Q$,
\begin{align}
    \Q\cdot \dl \psi=0,\quad  \dl \cdot (\rho\Q) =0, \quad \dl \times \lbr\B \times \Q  \rbr=0, \quad \dl \times \lbr \rho\u \times \Q \rbr=0.
    \label{Q_n_u_conditions}
\end{align}

The conditions on $\Q$ from \eqref{Q_n_u_conditions} reads
\begin{subequations}
\begin{align}
     \Q\cdot \dl \rho=0 \label{QDrho}\\
     \B \times \Q =\dl \psi \label{BxQ} \\
     \dl \cdot \Q =0,
     \label{divQ}
\end{align}
\label{Q_conditions}
\end{subequations}
while the conditions on $\u$ from \eqref{Q_n_u_conditions} and \eqref{ideal_MHD_flows} are
\begin{subequations}
\begin{align}
     \dl \cdot (\rho\u) =0 \label{div_rho_u}\\
     \u\times \B = \dl \Phi(\psi) \label{uxB}\\
     \dl \times \lbr \rho\u \times \Q \rbr=0.
     \label{curl_rho_uxQ}
\end{align}
\label{u_conditions}
\end{subequations}
We want to point out that the conditions given in \eqref{Q_conditions} are closely related to the ``weak quasisymmetry" conditions obtained in \citep{Rodriguez2020a}. If instead of $\QD\rho=0$, we assume $\QD B=0$ and $\rho=\rho(\psi,B)$ we get exactly the weak quasisymmetry (QS) conditions. This close connection of ideal MHD flows with quasisymmetry was pointed out earlier by Helander \citep{helander2007rapid,Helander2014}. 

The analogy with ``weak QS" becomes more evident when we look at the internal consistency of \eqref{Q_conditions}. For given $\B$ and $\rho$, the system \eqref{Q_conditions} is an overdetermined system for $\Q$ as it represents four constraints for the three components of $\Q$. From \eqref{BxQ} we get
\begin{align}
    \Q= \frac{1}{B^2}\lbr \BQ \B -\BxDpsi \rbr.
    \label{Q_form}
\end{align}
The component $\BQ$ can be chosen such that \eqref{QDrho} is satisfied. The divergence-free condition \eqref{divQ} then imposes the following condition on $\B$
\begin{align}
\J\cdot\dl\psi=\BD \BQ-\QD B^2,
    \label{J_del_psi}
\end{align}
which must be satisfied in order for \eqref{Q_conditions} to have a consistent solution. Such a consistency condition indeed appears in QS systems \citep{Rodriguez2020a,Burby2020} without the last term since $\QDB=0$.

If we assume that $\Q$ is given we can construct a $\B$ consistent with $\Q$. Using \eqref{BxQ}, we can write $\B$ in terms of $\Q$ as
\begin{align}
    \B=\frac{1}{Q^2}\lbr \BQ \Q +\Q\times\dl \psi \rbr,
    \label{BinQform}
\end{align}
which, is analogous to the symmetry flux coordinates in a tokamak \citep{haeseleerflux_coordinates} and the form used by \cite{Burby2020} to study the strong form of QS. To ensure that $\dl\cdot\B=0$ is satisfied by \eqref{BinQform}, we must have
\begin{align}
    \QD\BQ +\dl\cdot \lbr \Q\times\dl\psi \rbr -\BD Q^2=0.
    \label{divB_extra_condition}
\end{align}

In the following, we shall assume that there exists a Hameiri vector of the form $\C=\rho \Q$, where $\C\cdot \dl \rho=0$ and $\Q$ satisfies \eqref{Q_conditions} and \eqref{divB_extra_condition} and is given. 

\subsection{Underlying assumptions}
\label{subsec:basic_assumptions}
Before proceeding further, we shall discuss the various assumptions that we make in our analysis. We begin with our fundamental assumption that there exist non-symmetric plasma flows that possess nested flux surfaces. In our formal exploration, we are not forced to make any specific assumptions about the rotational transform of the magnetic field or the magnetic shear. The nestedness of flux surfaces is not guaranteed but is often made in the literature. In the following, we shall derive the necessary conditions (the Hamada conditions) required to suppress current singularities on the rational surfaces and preserve the nested surfaces. It will be apparent that the Hamada conditions can not possibly be satisfied in a toroidal volume unless magnetic shear is weak. With the analysis of previous sections in mind, we now make a conscious choice of weak magnetic shear in the following.

Our second important assumption is the existence of Hameiri's symmetry vector $\C$. Theoretically, $\C$ can be obtained by applying Noether's theorem to the one-fluid ideal MHD Lagrangian \citep{Ilgisonis_Pastukhov_2000_variational}. In particular, $\C$ is related to the relabelling symmetry of ideal MHD \citep{Ilgisonis_Pastukhov_2000_variational,hameiri1998variational,Vladimirov_Moffatt_1995_variational_MHD}. However, the system of equations determining $\C$, \eqref{C_conditions}, being a nonlinear overdetermined system, it is difficult to prove the existence of such a vector. Moreover, unlike QS, a systematic study of $\C$ has not been carried out to the best of our knowledge.  

As discussed in details in \citep{hameiri1998variational}, an important necessary condition that needs to be satisfied in order for $\C$ to exist in a toroidal domain \citep{hameiri1983_MHD_flow_equilibrium} is that on each closed magnetic field line,
\begin{align}
\oint \frac{d\ell}{B}\rho = m(\psi).
    \label{rho_condition}
\end{align}
Here, $m(\psi)$ is assumed to be a smooth function and $d\ell$ denotes the differential element along the  magnetic field. If condition \eqref{rho_condition} is not satisfied then the flow must be nearly-parallel \citep{hameiri1998variational}.

We then assumed that $\C\cdot\dl\rho=0$, which implies that the density (and hence pressure from \eqref{pressure_eqn} ) possess a continuous symmetry along $\C$. This choice has been made solely for simplicity and analytical tractability. Let us briefly discuss some of the consequences of the assumption of $\QD \rho=0=\QD p$. Firstly, we note that \eqref{rho_condition} is satisfied identically. Secondly, if $\Q$ lines do not close on themselves then density and pressure must not vary on a flux surface i.e.  $p=p(\psi),\rho=\rho(\psi)$. Therefore, such systems must be subsonic \citep{tasso1998axisymmetric} since large centrifugal forces lead to variations of density and pressure on flux surfaces. When $\Q$ lines are closed, flows do not have to be subsonic, and pressure variations on flux surfaces are allowed. Although we shall not make any assumption on the closure of $\Q$ lines, it is interesting to note that in QS, the symmetry lines close helically or toroidally depending on the type of QS.

The final assumption that $\Q$ satisfies \eqref{divB_extra_condition}, is due to the fact that \eqref{BinQform} is divergence-free only in axisymmetry  \citep{Burby2020}. The condition \eqref{divB_extra_condition} is not really an additional constraint on $\Q$ (or $\C=\rho \Q$), since it follows directly from \eqref{BxQ}. Any $\Q$ satisfying \eqref{BxQ} with a divergence free physical magnetic field necessarily satisfies \eqref{divB_extra_condition}. We have included the condition \eqref{divB_extra_condition} only because of its usage in subsequent calculations.

\subsection{ Generalized Bernoulli's law and angular momentum }\label{subsec:gen_Bernoulli}

It is well known \citep{hameiri1983_MHD_flow_equilibrium,Throumoulopoulos_tasso_Weitzner2006_nonexistence_purely_poloidal_flow} that five scalar functions are needed to describe the steady axisymmetric ideal MHD flows. Out of these, two functions are given by the electrostatic potential $\Phi'(\psi)$ and the entropy (assumed constant). The remaining three functions, denoted by $(\Lambda(\psi), H(\psi), I(\psi))$, are associated with the parallel component of the flow, the Bernoulli law, and angular momentum, respectively. In the following, our goal is to derive these quantities systematically for non-symmetric flows with the help of the Hameiri vector $\C=\rho \Q$. 

For a given magnetic field $\B$ and $\Q$ that satisfy \eqref{Q_conditions} and \eqref{J_del_psi}, it can be shown \citep{hameiri1998variational} that a steady state flow satisfying \eqref{u_conditions} is given by
\begin{align}
\u = \frac{\Lambda(\psi)}{\rho}\B - \Phi'(\psi) \Q,
    \label{u_form}
\end{align}
which clearly resembles the form of the axisymmetric steady flow of a tokamak \citep{hameiri1983_MHD_flow_equilibrium}. We also get the flux function $\Lambda(\psi)$ from \eqref{u_form}.

From \eqref{u_form}, we note that any two vectors from $\u,\B$ and $\Q$ can be used as basis vectors on a flux surface. We shall choose $\B$ and $\Q$ since the commutator of the two derivatives $\BD,\QD$ vanish (as shown in Appendix \ref{app:useful}). Choosing $\B,\Q,\dl\psi$ as basis vectors, we shall now obtain the components of the ideal MHD force balance equation \eqref{momentum_eqn}.

We find it convenient to rewrite \eqref{momentum_eqn} in the form of a vorticity equation
\begin{align}
    \dl\lbr \frac{1}{2}u^2 + h \rbr = \u \times \lbr \dl\times \u\rbr +\frac{1}{\rho}\JxB.
    \label{vorticity_eqn}
\end{align}
Here, we have defined $h$ such that $\dl h= (1/\rho)\dl p(\rho)$, and we have used the standard vector identity $\u\cdot\dl u = \dl (u^2/2) -\u\times (\dl\times \u)$. Dotting with $\B$ and $\Q$ respectively, and using (\eqref{uxB},\eqref{u_form}) we get
\begin{subequations}
\begin{align}
    \BD \lbr \frac{1}{2}u^2 + h \rbr + \dl\cdot \lbr \Phi'(\psi)\u \times \dl \psi \rbr=0 \label{temp_BD_vort}\\
   \rho \QD \lbr \frac{1}{2}u^2 + h \rbr - \J\cdot\dl\psi + \dl\cdot \lbr \Lambda(\psi) \u \times \dl \psi\rbr=0. \label{temp_QD_vort}
\end{align}
\label{temp_BD_QD_vort}
\end{subequations}
To evaluate the divergence terms that appear in \eqref{temp_BD_QD_vort} we use \eqref{uxB} and \eqref{uxdlpsi_n_its_div}. We note that the assumption $\QD \rho=0$ helps us to simplify \eqref{temp_QD_vort}. 

Simplifying \eqref{temp_BD_QD_vort} using \eqref{QDrho}, \eqref{J_del_psi} and \eqref{uxdlpsi_n_its_div} we get
\begin{subequations}
\begin{align}
    \BD \lbr \frac{1}{2}u^2 + h  + \Phi'(\psi)\uQ \rbr -\QD \lbr \uB \Phi'(\psi)\rbr=0 \label{BD_vort}\\
   \BD \lbr \lbr \B -\Lambda(\psi)\u \rbr\cdot \Q\rbr- \QD \lbr \frac{1}{2}\rho u^2 +B^2 -\Lambda(\psi) \uB \rbr =0. \label{QD_vort}
\end{align}
\label{BD_QD_vort}
\end{subequations}
In axisymmetry, the $\QD$ terms in \eqref{BD_QD_vort} vanish identically, and we get two homogeneous magnetic differential equations. Solving the homogeneous magnetic differential equations, we get two flux functions which denote Bernoulli's law and momentum conservation in the symmetry ($\Q$) direction \citep{hameiri1983_MHD_flow_equilibrium,tasso1998axisymmetric}. The situation is similar here except that the magnetic differential equations are non-homogeneous; therefore, we need to check for consistency conditions. We postpone the discussion on the consistency conditions until section \ref{subsec:last_Hamada}. 

Assuming that the solvability conditions are satisfied, we solve \eqref{BD_QD_vort} and obtain the generalized Bernoulli and the generalized angular momentum equations
\begin{subequations}
\begin{align}
     &\frac{1}{2}u^2 + h  + \Phi'(\psi)\uQ =H(\psi) +\cH\quad \quad \quad \text{(Bernoulli)} \label{Bernoulli}
     \\
    &\BQ-\Lambda(\psi)\uQ = I(\psi)+\cI,\quad \quad\text{(angular momentum)}
     \label{Q_momentum}
\end{align}
\label{Bernouili_n_Q_momentum}
\end{subequations}
where, $\cH,\cI$ defined by
\begin{align}
    \cH = \QD \int\frac{d\ell}{B} \uB \Phi'(\psi), \quad  \cI =\QD\int \frac{d\ell}{B}\lbr \frac{1}{2}\rho u^2 +B^2 -\Lambda(\psi) \uB \rbr,
    \label{cH_n_cI}
\end{align}
are single-valued and doubly-periodic functions if Hamada conditions (see  \eqref{solvability_u_B_circulations} below) are satisfied. They vanish identically in the axisymmetric case. As is well-known \citep{beskin2009mhd,hameiri1998variational,hameiri1983_MHD_flow_equilibrium} availability of conserved quantities are very helpful in characterizing flow patterns. 

\subsection{ Generalized Grad-Shafranov equation for flows with the Hameiri vector }\label{subsec:GGS_flow}

We are now in a position to obtain the GGS from the $\dl \psi \cdot $ component of the vorticity equation \eqref{vorticity_eqn}. The step-by-step derivation is provided in Appendix \ref{app:GGS}. The steps to derive the GGS are identical to the ones in the derivation of the flow-modified classical Grad-Shafranov equation in axisymmetry or helical symmetry. However, there are important differences which we shall now discuss. Firstly, in the absence of axial/helical symmetry, the vector
\begin{align}
    \w= \lbr \dl \times \Q \rbr \times \Q +\dl Q^2,
    \label{w_form}
\end{align}
which denotes the deviation of $\Q$ from axial/helicalsymmetry vector \citep{Burby2020} appears in the GGS. We can check that \eqref{divB_extra_condition}, the condition required to ensure $\dl\cdot\B=0$ when $\B$ is expressed in terms of $\Q$ (equation \eqref{BinQform}), is identical to
\begin{align}
    \QD \BQ - \B\cdot \w=0.
\end{align}
Obviously, if $\Q$ is an axial/helical symmetry both terms individually vanish and $\B$ is automatically divergence-free. Secondly, the geometric quantity 
\begin{align}
    \omega_Q\equiv\Q\cdot \dl \times \Q,
    \label{QdotcurlQ}
\end{align} 
which doesn't appear in the axisymmetric GS, but appears in the helically symmetric GS, also appear in the GGS. Finally, terms appear with the force-like quantity 
\begin{align}
    \bm{F_Q}= \J\times \Q + \dl \BQ,
    \label{FQ_form}
\end{align}
which vanishes identically in static MHD \citep{Burby2020}. 

With the definitions (\eqref{w_form},\eqref{QdotcurlQ} and \eqref{FQ_form}) in mind, we can write the flow-modified GGS in the following form 
    \begin{align}
     \dl\cdot \lbr\lbr 1-\frac{\Lambda^2}{\rho}\rbr \frac{1 }{Q^2} \dl \psi\rbr +\rho \frac{dH}{d\psi}+&\uB \frac{d\Lambda}{d\psi} + \frac{\BQ}{Q^2}\frac{dI}{d\psi} - \rho\uQ \frac{d\Phi'}{d\psi} +\cN=0,  \label{GGS}
     \end{align}
 where,    
    \begin{align}
    \cN=\frac{I+\cI}{Q^2}\lbr \frac{\omega_Q}{Q^2} + \frac{\bm{F_Q}\cdot \dl\psi}{|\dl\psi|^2}\rbr&+ \lbr \rho \dl \cH + \frac{\BQ}{Q^2}\dl \cI\rbr\cdot \frac{\dl\psi}{|\dl\psi|^2}
   \nonumber \\
    &+\frac{\w\cdot \dl \psi}{Q^2} \lbr \frac{1}{Q^2}\lbr 1-\frac{\Lambda^2}{\rho} \rbr - \frac{\rho \Phi'\uQ}{|\dl\psi|^2}\rbr. \label{cN} 
\end{align}
In axisymmetry \citep{hameiri1983_MHD_flow_equilibrium}, $\cN=0$. For helical symmetry, an additional term of the form $(I\omega_Q)/Q^4$ appears. To study force-balance other than ideal MHD force-balance, we can replace $\bm{F_Q}$ by other forces \citep{Rodriguez_Sengupta_Bhattacharjee_QS_flow}. For the case of flows only in the symmetry direction $\Lambda(\psi)=0$ and for strictly parallel flows $\Phi'(\psi)=0$. One can easily extend the GGS to include other non-relativistic forces.

Unlike in symmetric geometries, the coefficients and derivatives in the flow-modified GGS depend on all three spatial variables \citep{Constantin2020,Burby2020}. Hence, we need to impose the additional condition 
\begin{align}
    \Q \cdot \dl \psi=0.
    \label{Qdelpsi_s}
\end{align}
Whether such solutions to the GGS can be constructed is a challenging open problem.

\subsection{ The generalized Hamada conditions }\label{subsec:last_Hamada}
We now return to the consistency conditions that must be satisfied in order for \eqref{BD_QD_vort} to have non-singular smooth solutions. When the rotational transform is irrational, we can flux-surface average \eqref{BD_QD_vort}. Since both $\B$ and $\C$ are tangential to the flux surfaces, there is no inconsistency. On the other hand, for rational rotational transform, the vanishing of the two equations under closed field line $\oint d\ell/B$ integral is not automatic and leads to nontrivial constraints \citep{Newcomb_1959_MDE}- the so-called \textit{Hamada conditions} \citep{Hamada1962_coordinates,Helander2014}. 

We recall that in ideal MHD equilibrium with scalar pressure $p(\psi)$, the integral constraints that must be satisfied on rational surfaces are \citep{Grad1967toroidal_confinement,Soloviev_Shafranov_v5_1970plasma,Grad1971plasma}
\begin{align}
    \oint \frac{d\ell}{B} = c_1(\psi), \quad \oint  \frac{d\ell}{B} B^2= \oint \B\cdot \bm{d\ell} = c_2(\psi), \quad c'_2(\psi) + p'(\psi) c_1(\psi)=0,
    \label{ideal_MHD_dlB_oonstraints}
\end{align}
where, $c_1(\psi),c_2(\psi)$ are single-valued continuous functions of $\psi$ that reduce to constants in the vacuum limit. We give a straightforward proof of these conditions in Appendix \ref{app:consistency_conditions}. 

Hamada conditions \eqref{ideal_MHD_dlB_oonstraints} are identically satisfied in symmetric geometries but are not satisfied in general for non-symmetric geometry with arbitrary pressure and rotational transform profiles \citep{Boozer_coords_1981,Weitzner2016}, leading to singular currents on rational surfaces  \citep{Loizu_2015_existence_current_sheets,Loizu_2015_magnetic_islands_singular_currents}. In the following, we shall obtain the generalized Hamada conditions for steady flows with Hameiri's vector $\C=\rho\Q$. 

We note that \eqref{rho_condition} is already in the form of a Hamada condition. To find the other conditions, we carry out the $\oint d\ell/B$ integrals of \eqref{BD_QD_vort} (denoted by $\langle \quad \rangle$ brackets ). Using the commutation of the two derivatives to get two homogeneous equations of the form $\QD \langle  A \rangle =0$, which implies that $A$ must be a flux function. Thus, we obtain two consistency conditions
\begin{subequations}
\begin{align}
\langle\u \cdot\B \rangle&= \oint \u\cdot \bm{d\ell}=C_1(\psi)
\label{solvability_u_circulations}\\ 
\bigg\langle B^2 + \frac{1}{2}\rho u^2 -\Lambda \uB\bigg\rangle &= \oint \B\cdot \bm{d\ell}+\oint \frac{d\ell}{B}\lbr \frac{1}{2}\rho u^2  \rbr -\Lambda(\psi) C_1(\psi) =C_2(\psi),
 \label{solvability_B_circulations}
\end{align}
\label{solvability_u_B_circulations}
\end{subequations}
which show that the circulations of $\u$ and $\B$ along the closed field line are required to be flux functions in order that \eqref{BD_QD_vort} be solvable.

The final Hamada condition is obtained by closed field line averaging of \eqref{GGS} on a rational flux surface, which leads to the following equation for $C_2'(\psi)$ (see Apendix \ref{app:derivation_final_Hamada})
\begin{align}
C'_2(\psi)+C_1(\psi)\frac{d\Lambda}{d\psi}= H\del_\psi \langle\rho\rangle  -\Phi'\del_\psi \langle\rho\uQ \rangle  -\langle \cN \rangle + \del_\psi\bigg\langle \rho \cH +\frac{\cI}{Q^2}\BQ \bigg\rangle.
    \label{C2prime_comstraint}
\end{align}

Equations \eqref{solvability_u_B_circulations} and \eqref{C2prime_comstraint}, together with \eqref{rho_condition} describe integral constraints that need to be satisfied on each rational flux surface. These are the generalized Hamada conditions that include the effect of steady flows. In Appendix \ref{app:consistency_conditions} we show that when these consistency conditions are satisfied magnetic resonances on rational surfaces are avoided. Given the complicated nature of the generalized Hamada conditions, it is doubtful that they will be satisfied by a generic steady flow. Furthermore, since the Hamada conditions need to be satisfied on each rational surface, the magnetic shear needs to be sufficiently weak to avoid passage through multiple low-order rational surfaces.

\subsection{Summary of section \ref{sec:GGS_C} }
\label{subsec:summary}
In section \ref{sec:GGS_C}, we discussed a special class of non-symmetric steady flows endowed with Hameiri's symmetry vector $\C=\rho \Q$, where $\C\cdot \dl \rho =0$ and $\Q$ satisfy the overdetermined system \eqref{Q_conditions}. If the integral constraints \eqref{rho_condition}, \eqref{solvability_u_B_circulations} and \eqref{C2prime_comstraint}, are satisfied then the description of such flows involve a GGS \eqref{GGS} and four flux-functions similar to axisymmetric flows \citep{hameiri1983_MHD_flow_equilibrium, tasso1998axisymmetric} namely, $\Lambda(\psi),\Phi'(\psi),H(\psi),I(\psi)$. Note that only the first four appear in the GGS since we have assumed barotropic equation of state with constant entropy. The flux function $m(\psi)$ is needed to ensure existence of Hameiri's symmetry vector. The integral constraints \eqref{solvability_u_B_circulations} are necessary to avoid current singularities (see Appendix \ref{app:consistency_conditions}). They furnish two more flux-functions $C_1(\psi), C_2(\psi)$ which measure the circulations of $\u$ and $\B$ in a closed field line system.  Physically, they amount to the constraint that the flows should not lead to accumulation of charges inside the closed magnetic flux tubes on rational surfaces. Finally, non-symmetric geometry introduces several extra terms grouped together as $\cN$ such as $\cH,\cI, \w\cdot \dl \psi$ etc.  

\section{Discussion}\label{sec:discussion}
We studied several classes of non-symmetric ideal MHD equilibrium with flows larger than the diamagnetic flows. Such flows could occur in stellarators with We showed that the techniques developed in carrying out perturbation methods to all orders for static MHD equilibrium \cite{weitzner2014ideal} can also be extended to MHD equilibrium with flows. The basic idea is that if the field lines are closed, i.e., the rotational transform is rational and the magnetic shear is weak, one can systematically eliminate magnetic resonances at each order by utilizing the ``free-function" that one gets by solving the magnetic differential equation from lower orders. Plasma flows introduce extra resonances, which are absent in static MHD. It is, in general, very complicated to eliminate all such resonances. However, for the class of equilibrium with nearly parallel flows, one can do so. It is to be noted that exact solutions with large parallel flows can be constructed \citep{kamchatnov1982topological_soliton_mhd} when the flow is considered incompressible. We have argued that compressible, nearly parallel flows are also possible in the MHD model of plasma.

Although the present work on the analysis of magnetic resonances has been carried out on rectangular coordinates with periodic boundary conditions, extensions to polar and toroidal coordinates should be fairly direct. The principal issue is behavior near the magnetic axis. To deal with this complication, one must require that components of the fields and flows vanish sufficiently rapidly near the axis. The interactions through the nonlinear terns typically do not destroy these properties. We leave the detailed near-axis analysis for the future.

We have studied a particular class of perpendicular flows by taking a more formal approach. Our approach is based on utilizing a symmetry vector found by Hameiri \citep{hameiri1981spectral_C,hameiri1998variational}. The generalizations of Bernoulli's law and conservation of angular momentum were also obtained thanks to Hameiri's symmetry vector. We obtain a flow-modified generalized Grad-Shafranov equation for such flows and the constraints resulting from the closed magnetic field lines. It is clear from the nature of the constraints that if such non-parallel flows exist, they must be special. There exists a close connection between Hameiri's vector and the quasisymmetry vector that will be further discussed elsewhere \citep{Rodriguez_Sengupta_Bhattacharjee_QS_flow}.

We have not attempted to estimate the damping of the flows due to neoclassical effects, which is an essential but difficult question \citep{Simakov_Helander_2009_neoclassical_mom_transp}. We shall address this in detail in a forthcoming paper \citep{Rodriguez_Sengupta_Bhattacharjee_QS_flow}. We only make a few observations here. First, the neoclassical effects are tied to the parallel electric field, which does not appear in ideal MHD. If the parallel electric field can be shown to be self-consistently small or higher-order in the Larmor radius, then the MHD description prevails, and flow damping would be minimized. Therefore, our results are valid within the framework of ideal MHD. Second, the techniques developed here can be easily extended to more sophisticated MHD models that are relevant to astrophysical plasmas.

Our results here do not necessarily contradict earlier results \citep{Simakov_Helander_2009_neoclassical_mom_transp, sugama2011_QS_large_flow} that argue that large flows can not be supported even if the stellarator is quasisymmetric but not completely axisymmetric. We show that it is challenging to avoid magnetic resonances unless the flow is nearly parallel. Except for very special flows, the generalized Hamada conditions on rational surfaces can not be satisfied, severely limiting the possibility of steady non-symmetric flows in generic stellarators. We expect that kinetic constraints will further restrict the class of available solutions.

\section*{ Acknowledgements}
The authors would like to sincerely thank the anonymous reviewer for their excellent, meticulous, and constructive suggestions that have significantly improved the organization and presentation of the results in the paper. In addition, WS would like to thank E. Rodriguez, A.Bhattacharjee, A.B.Hassam, E.J.Paul, and J.Juno  for helpful discussions. This research was partly
funded by the US DOE grant no. DEFG02-86ER53223 and Simons Foundation/SFARI (560651, AB).

\appendix

\section{Analysis of the dispersion relation \eqref{The_disp}}
\label{app:The_disp}
Our goal here is to show that real solutions of the dispersion relation \eqref{The_disp} exist. We shall rewrite the dispersion relation in terms of dimensionless physical quantities and analyze the parametric space of these parameters, which support real solutions of \eqref{The_disp}. We shall also look at some simple, special cases which will allow us to gain insight into the parametric space.

Let us consider a generic flow of the form
\begin{align}
    \bm{u}=u_\parallel \hat{\B}+\upr , \quad  \hat{\B}= \frac{\B}{B}, \quad u_\parallel = \bm{u}\cdot \hat{\B}.
\end{align}
Associated with this flow, we can define the parallel and perpendicular Mach numbers as follows
\begin{align}
    M_\parallel= \frac{u_\parallel}{c}=\frac{\bm{u}\cdot \B}{c B}, \quad M_\perp = \frac{u_\perp}{c}= \frac{|\bm{u}_\perp|}{c},
    \label{Mach_def}
\end{align}
$c$ being the sound speed.

As noted in the text, the dispersion relation is independent of the magnitude of $\bm{k}$. Therefore, we can divide \eqref{The_disp} by $(\rho c^2 k^2)^2$ to get
\begin{align}
    (k_u)^4 -(k_u)^2\left(1+\frac{1}{\beta}\right) + (k_B)^2 \frac{1}{\beta}=0,
    \label{The_disp_reln}
\end{align}
where,
\begin{align}
    \frac{1}{\beta}= \frac{B^2}{\rho c^2},\quad  k_u = \bm{\hat{k}}\cdot \frac{\u}{c}, \quad  k_B = \bm{\hat{k}}\cdot \hat{\B}.
    \label{beta_kB_ku_def}
\end{align}
We will not consider any components of $\bm{\hat{k}}$ orthogonal to $\bm{u}$ and $\B$ since they trivially satisfy \eqref{The_disp_reln}. Hence, we will assume that the unit vector $\hat{k}$ is coplanar with $\bm{u},\B$ and therefore, admits the following orthogonal decomposition
\begin{align}
    \bm{\hat{k}}=  \cos{\theta}\:  \hat{\B} + \sin\theta\: \hat{\bm{u}}_\perp, \quad \hat{\bm{u}}_\perp= \frac{\bm{u}_\perp}{u_\perp}.
\end{align}
Furthermore, \eqref{beta_kB_ku_def} together with \eqref{Mach_def} implies that
\begin{align}
    k_B= \cos\theta, \quad k_u = M_\perp \sin{\theta} + M_\parallel \cos{\theta}.
    \label{kB_ku_theta_form}
\end{align}
To determine $\theta$ we substitute \eqref{kB_ku_theta_form} into the dispersion relation \eqref{The_disp_reln}. which leads to the following quartic equation for $\xi =\tan\theta$

\begin{align}
   &a_4 \xi^4 +a_3 \xi^3 + a_2 \xi^2 +a_1 \xi +a_0=0, \label{The_xi_quartic}\\
   &a_4= M_\perp^2 \left(M_\perp^2 -\left(1+\frac{1}{\beta}\right)\right), \quad \quad a_0= M_\parallel^2 \left(M_\parallel^2 -\left(1+\frac{1}{\beta}\right)\right)+\frac{1}{\beta}, \nonumber\\
    & a_3=4 M_\parallel M_\perp\left(M_\perp^2 -\frac{1}{2}\left(1+\frac{1}{\beta}\right)\right), \quad  a_1=4 M_\parallel M_\perp\left(M_\parallel^2 -\frac{1}{2}\left(1+\frac{1}{\beta}\right)\right),
   \nonumber\\
   &a_2=6 \left(M_\parallel M_\perp \right)^2-\left(1+\frac{1}{\beta}\right)\left(M_\parallel^2+ M_\perp^2\right)+\frac{1}{\beta} \nonumber.
\end{align}
We note that there are only three physical parameters that come into the problem: plasma beta, $\beta=\rho c^2/B^2$, parallel and perpendicular Mach numbers $(M_\parallel, M_\perp)$ given by \eqref{Mach_def}. In the following, we shall assume that all three physical quantities are real and positive.

\begin{figure}
     \centering
     \begin{subfigure}[t]{0.33\textwidth}
         \centering
         \includegraphics[width=\textwidth]{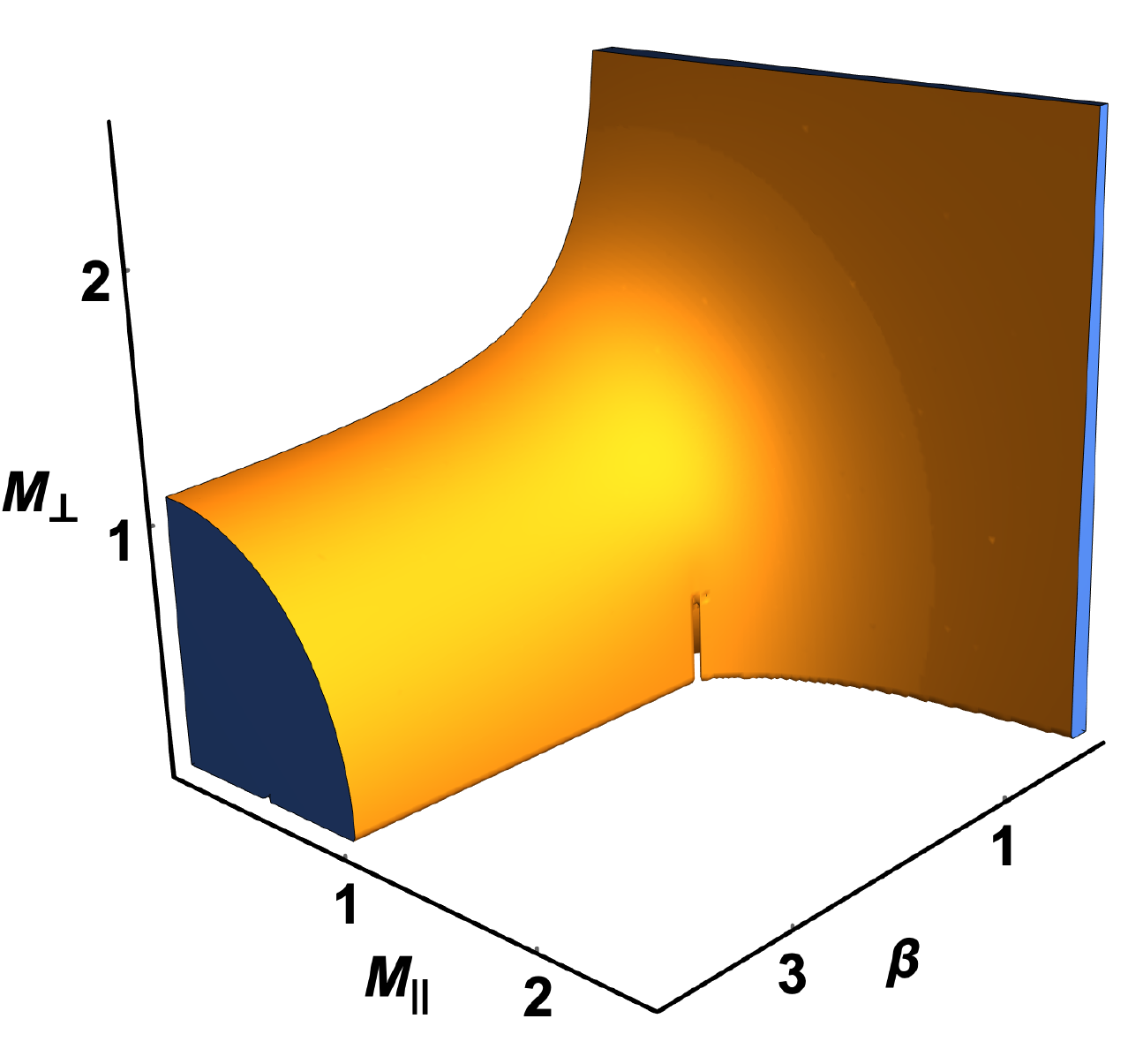}
         \caption{region with 2 real roots}
         \label{fig:2_real_roots}
     \end{subfigure}
     \hfill
     \begin{subfigure}[t]{0.33\textwidth}
         \centering
         \includegraphics[width=\textwidth]{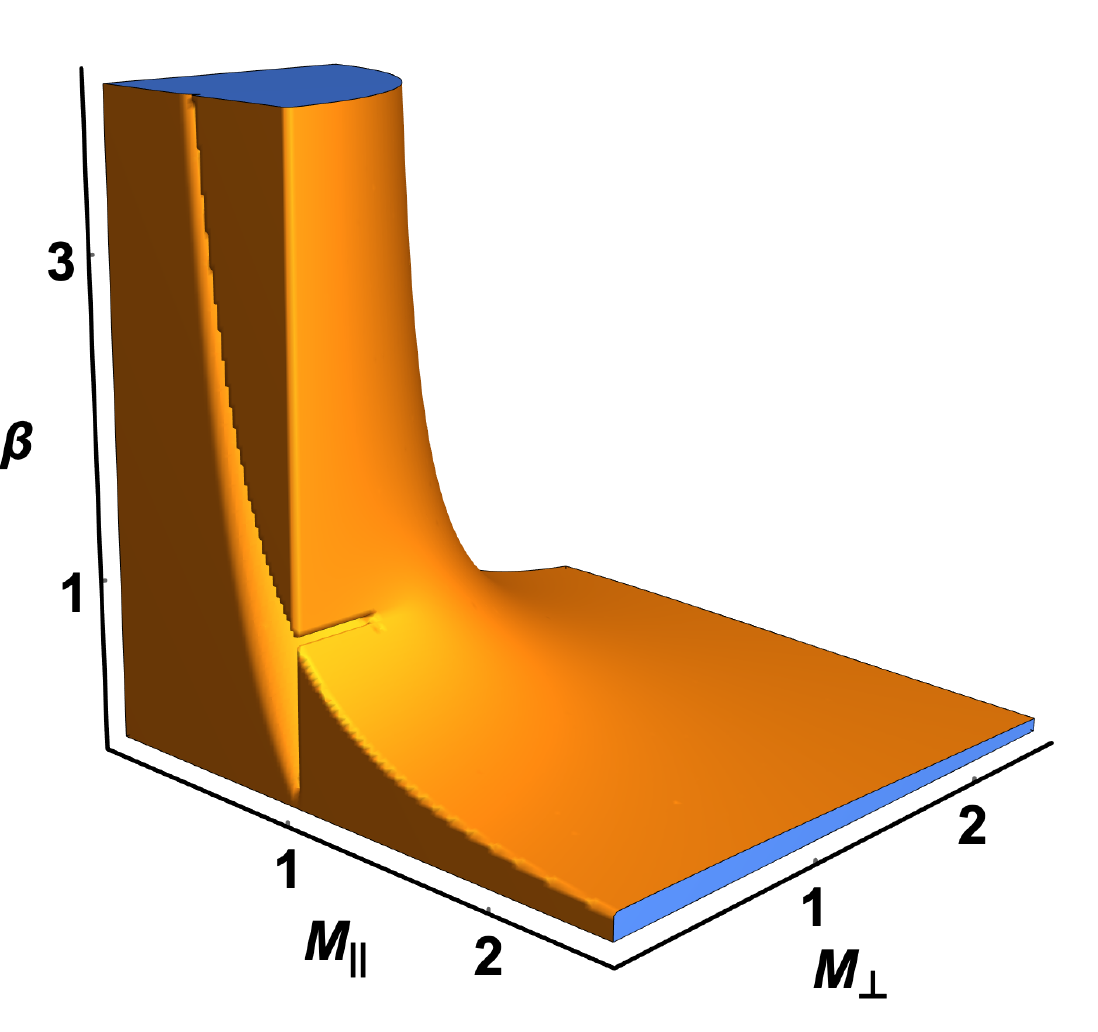}
         \caption{side-view }
         \label{fig:2_real_roots_side}
     \end{subfigure}
     \hfill
     \begin{subfigure}[t]{0.3\textwidth}
         \centering
         \includegraphics[width=\textwidth]{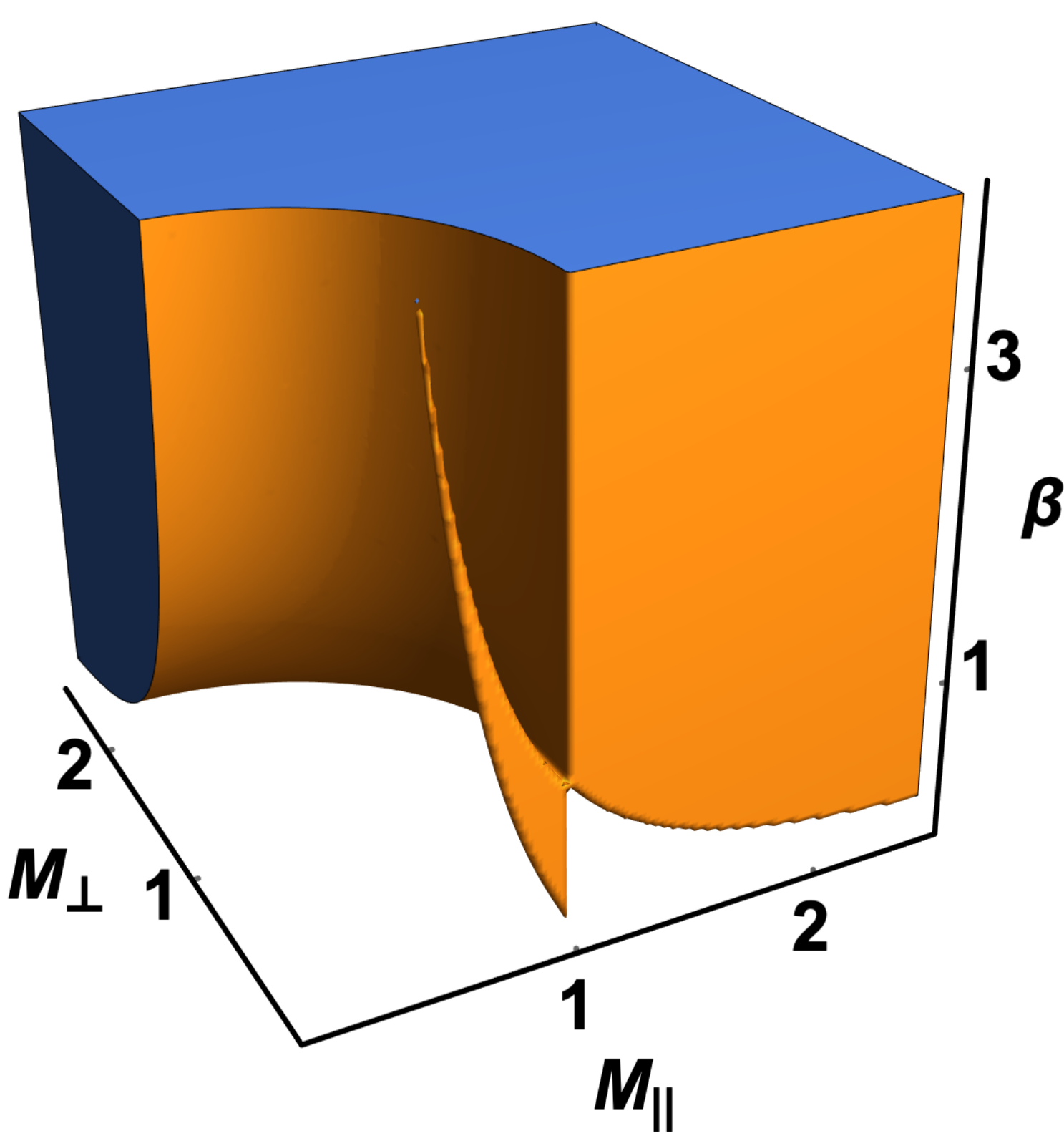}
         \caption{region with 4 real roots}
         \label{fig:4_real_roots}
     \end{subfigure}
        \caption{a) and b) The region of parameter space where \eqref{The_xi_quartic} has 2 real and 2 imaginary roots as viewed from two sides. c) The region with four real roots. Since the latter is complimentary to the former, the equation has at least one real root. Along the lines $M^2_\parallel=1,1/\beta, 1/(\beta+1)$ and $\beta=1$ multiple roots co-exist (see text). }
        \label{fig:three graphs}
\end{figure}

The theory of quartic equations is well-developed, and the necessary conditions for a quartic equation to have real roots can be determined straightforwardly \citep{Rees_1922_quartic_eqn}. When the quartic discriminant is negative, the equation has two real and two imaginary roots. We have plotted the associated parametric region in Figure \eqref{fig:2_real_roots} and \eqref{fig:2_real_roots_side}. The region where four real roots can exist is shown in Figure \eqref{fig:4_real_roots}. These two regions are complimentary and hence the equation always has a real root. The details are not particularly illuminating and will be omitted here. Fortunately, a lot of insight can be gained by studying some simple cases, which we shall do next.

We begin with the special case of purely parallel flows i.e. $M_\perp=0$. In this case, the equation simplifies considerably and we get
\begin{align}
    \xi^2= \left(\frac{M_\parallel^2-\frac{1}{\beta}}{\frac{1}{\beta +1}-M_\parallel^2}\right)\frac{\left(1-M_\parallel^2\right)}{\left(1+\frac{1}{\beta}\right)}.
\end{align}
As $M_\parallel^2\to 1/(\beta+1), \xi \to \infty$, and the angle $\theta=\arctan{\xi}\to \pi/2$, which implies that $\bm{k}$ becomes orthogonal to $\hat{B}$. On the other extreme, when $M_\parallel^2=1$ or $M_\parallel^2=1/\beta$, $\xi=0$, i.e. $\bm{k}$ is parallel to $\hat{B}$. The case $M_\parallel^2=1/\beta$ is discussed in \citep{kamchatnov1982topological_soliton_mhd}. These special solutions appear as lines of discontinuity in Figure \eqref{fig:2_real_roots_side} and \eqref{fig:4_real_roots}. The quartic determinant is zero along these curves leading to multiple roots. The solutions corresponding to $M_\parallel^2=1/(\beta+1),1/\beta$, for flows of the type \eqref{29} were described as pathological solutions in \eqref{30} which lead to $\Delta=0$. Physically, the system can develop shocks along these curves. 

Omitting the three special cases discussed above, we find that real solutions with $\xi>0$ exist for $M_\parallel$ less (greater) than one, provided $M_\parallel^2$ lies inside (outside) of the interval $(1/(\beta +1),1/\beta)$.  We note that for large values of $\beta$ subsonic parallel ($M_\parallel^2<1$) flows are rare compared to the  supersonic parallel flows, in accordance with Figure \eqref{fig:2_real_roots}.

Next, we consider the case of purely perpendicular flows, i.e., $M_\parallel=0$. If
\begin{align}
    M_\parallel=0 \quad \text{and} \quad M_\perp^2= 1+\frac{1}{\beta} \quad , \quad  \xi^2=\frac{1}{1+\beta +\frac{1}{\beta}}.
    \label{special_case_Mp0}
\end{align}
Since $\xi^2$ is positive definite, $\xi$ is real and the flow is perpendicular and supersonic. 

If $M_\parallel=0$ but $M_\perp$ does not satisfy \eqref{special_case_Mp0}, equation \eqref{The_xi_quartic} reduces to the following bi-quadratic equation
\begin{align}
    \xi^4 + \frac{(1+\beta)}{M_\perp^2 \beta}\left(\frac{\frac{1}{\beta+1}-M_\perp^2}{M_\perp^2-\frac{1+\beta}{\beta}}\right)\xi^2 +\frac{1}{M_\perp^2 \beta}\frac{1}{M_\perp^2-\frac{1+\beta}{\beta}}=0.
    \label{bi_quadratic}
\end{align}
For the roots to be real, the discriminant of \eqref{bi_quadratic} must be  positive, which implies 
\begin{align}
    M_\perp^2 > \frac{\frac{\beta+1}{\beta-1}+2\sqrt{\beta}}{\beta-1} \:\:\text{or}\:\: M_\perp^2 < \frac{\frac{\beta+1}{\beta-1}-2\sqrt{\beta}}{\beta-1}.
    \label{discriminant_Mperp}
\end{align}
We find that $\beta=1$ represents a discontinuity in the solution as depicted in Figure \eqref{fig:2_real_roots} and \eqref{fig:2_real_roots_side}. For values of $\beta$ close to zero we find that perpendicular flows can exist only if $M_\perp \approx 1$. For large $\beta$ we find that the first condition is easily satisfied but not the second one. 

Furthermore, for $\xi$ to have four real roots we need
\begin{align}
    M_\perp^2 > \left(1+\frac{1}{\beta}\right) \:\:\& \;\:  M_\perp^2 >\frac{1}{\beta+1}.
\end{align}
On the other hand, when $M_\perp^2<1+1/\beta$,  $\xi$ has at least two real roots provided \eqref{discriminant_Mperp} is satisfied. Therefore, both supersonic and subsonic perpendicular flows can exist.

\begin{figure}
    \centering
    \includegraphics[width=0.4\textwidth]{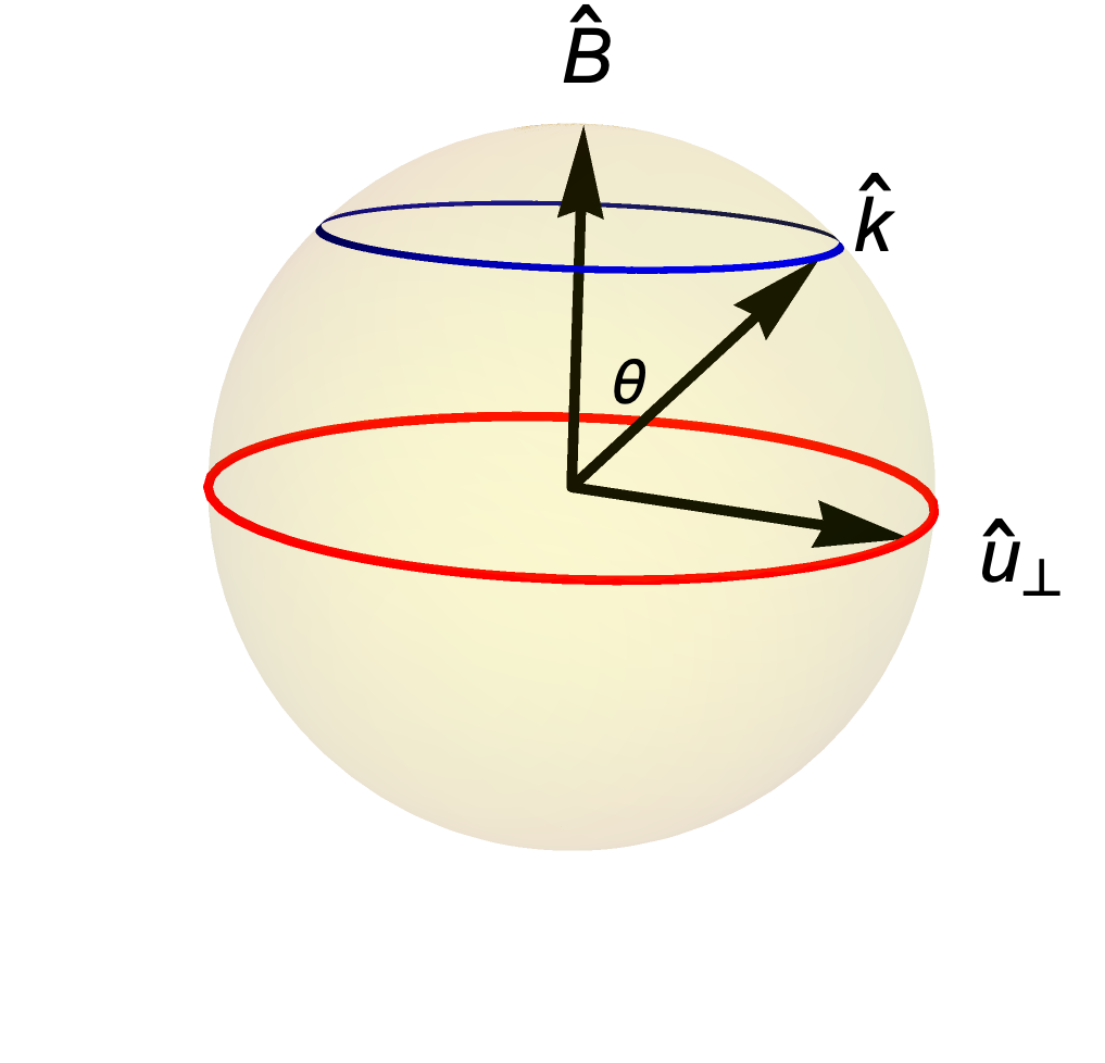}
    \caption{Depiction of the dispersion relation \eqref{The_disp_reln} as a curve on a unit sphere with coplanar unit vectors $(\bm{\hat{B}},\bm{\hat{k}},\bm{\hat{u}_\perp})$. As the unit vector $\hat{u}_\perp$ moves on the red circle, $\bm{\hat{k}}$ traces out the blue circle. $\xi=\tan{\theta}$ satisfies \eqref{The_xi_quartic}. }
    \label{fig:uBk}
\end{figure}

Finally, we note that $\bm{\hat{k}},\hat{\B}, \bm{\hat{u}_\perp}$ can be plotted on a unit sphere as shown in Figure \ref{fig:uBk}. We choose $\hat{\B}$ to be in the $z$ direction and, therefore, $\bm{\hat{u}_\perp}$ must lie in the $x-y$ plane. The coplanar vector $\bm{\hat{k}}$ makes an angle $\theta$ with $\B/B$, where $\theta =\arctan{\xi}$. Since the equation for $\xi$, \eqref{The_xi_quartic}, depends on $\bm{u}$ only through the combinations $M_\parallel \propto \bm{u}\cdot \B$ and $M_\perp \propto |\bm{u}_\perp|$, it is independent of the direction of $\hat{\bm{u}}_\perp$. As a result, the curve that $\bm{k}$ traces out is symmetric around $\hat{B}$ and hence a circle on the unit sphere.

\section{Useful identities and expressions}
\label{app:useful}
We collect some useful identities and expressions in this appendix that have been used in deriving some of the key results in the text.

A useful expression is that of the commutator $ [\BD,\QD]$. Employing standard Einstein summation convention, $\BD=B_i\del_i$, $\QD=Q_j\del_j$, together with \eqref{uxB} implies that $\ep_{klm}B_l Q_m = \del_k \psi$. It is straight-forward to show that  
\begin{align}
    [\BD,\QD]f=\BD (\QD f)-\QD(\BD f)=\dl\cdot \lbr \dl \psi \times \dl f\rbr =0
    \label{commutator}
\end{align}

The following identity is useful to simplify the divergence of $ \u\times\dl\psi$
\begin{subequations}
\begin{align}
    \u\times\dl\psi = \u \times (\B\times \Q)= \uQ \B -\uB \Q
    \\
    \dl\cdot \lbr \u\times\dl\psi\rbr = \BD \uQ -\QD\uB.
\end{align}
\label{uxdlpsi_n_its_div}
\end{subequations}

\section{Closed field line consistency conditions and current singularities }
\label{app:consistency_conditions}
The close relation between current singularities that can develop on rational surfaces in a nonsymmetric toroidal domain in ideal MHD, and the necessary conditions in the form of integrals over closed magnetic field lines such as $\oint d\ell/B = f(\psi)$ are well known \citep{Hamada1962_coordinates,Grad1967toroidal_confinement,Soloviev_Shafranov_v5_1970plasma,Boozer_coords_1981,Loizu_2015_magnetic_islands_singular_currents}. We show here that to avoid current singularities on rational surface the integral constraints of the form \eqref{ideal_MHD_dlB_oonstraints} for ideal MHD equilibrium, and \eqref{solvability_u_B_circulations}, \eqref{C2prime_comstraint} for steady flows, are sufficient conditions.

We shall use a generic $(\psi,\alpha,\varphi)$ coordinate system, where $\psi$ is the flux-label, $\alpha$ is the field-line label, $\varphi$ denotes another angle like coordinate. We will assume that the magnetic field lines are closed and shear is weak such that both $\dl \alpha$ and $\dl \varphi$ are single-valued. A more general derivation for finite shear can be carried out using generalized Boozer coordinates.

In the $(\psi,\alpha,\varphi)$ coordinates, we can represent $\B$ in the following covariant and contravariant forms
\begin{align}
\B= B_\psi \dl \psi + B_\alpha \dl \alpha + B_\varphi \dl \varphi,\quad \B = \dl \psi\times \dl \alpha.
    \label{Bcon_Bcov_forms}
\end{align}
The functions $B_\psi,B_\alpha, B_\varphi$ are all single-valued functions since $\B$ is single-valued and so are the gradients of $(\psi,\alpha,\varphi)$. The Jacobian $\cJ^{-1}= \dl\psi\times \dl \alpha\cdot \dl \varphi= \BD \varphi$ connects $B_\varphi$ to $B^2$ through
\begin{align}
    B_\varphi = \cJ B^2. \label{B_varphi_exp}
\end{align}
The current obtained by taking the curl of the covariant form of $\B$ satisfies
\begin{align}
    \J = \dl B_\psi \times \dl \psi + \dl B_\alpha \times \dl \alpha + \dl B_\varphi \times \dl \varphi\\
    \JxB = \dl \psi \lbr \BD B_\psi -\cJ^{-1}\del_\psi B_\varphi \rbr + \dl \alpha \lbr \BD B_\alpha -\cJ^{-1}\del_\alpha B_\varphi \rbr
    \label{J_and_JxB}
\end{align}
We now consider a general force-balance equation of the form
\begin{align}
\JxB = \F, \quad \F = F_\psi \dl \psi +F_\alpha \dl \alpha.
    \label{gen_force_bal}
\end{align}
Note that $\F$ does not have a $\dl \varphi$ component because of parallel force balance condition $\F\cdot \B=0$. 

If the current $\J$ simultaneously satisfies $\J=\dl\times \B$ and $\JxB=\F$, we can equate $\JxB$ from \eqref{gen_force_bal} to \eqref{J_and_JxB} to get the following magnetic differential equations for $B_\psi$ and $B_\alpha$ that enforce force-balance
\begin{align}
    \BD B_\psi -\cJ^{-1}\del_\psi B_\varphi = F_\psi, \quad \BD B_\alpha -\cJ^{-1}\del_\alpha B_\varphi = F_\alpha.
    \label{MDES_Bpsi_Balpha}
\end{align}
Eliminating $B_\varphi$ we obtain another magnetic differential equation
\begin{align}
    \BD\lbr \del_\alpha B_\psi-\del_\psi B_\alpha\rbr =\cJ^{-1}\lbr \del_\alpha (\cJ F_\psi)- \del_\psi (\cJ F_\alpha) \rbr
    \label{MDE_the_third}
\end{align}
Since $B_\psi$ and $B_\alpha$ are single-valued, the well-known necessary and sufficient conditions for the solvability of magnetic differential equations \citep{Newcomb_1959_MDE} leads to the following constraints
\begin{align}
    \del_\psi \oint d\varphi B_\phi + \oint d\phi \cJ F_\psi=0, \quad \del_\alpha \oint d\varphi B_\phi + \oint d\phi \cJ F_\alpha=0
    \label{temp_Newcomb}\\
    \del_\alpha \oint d\phi \cJ F_\psi - \del_\psi \oint d\phi \cJ F_\alpha=0
    \nonumber 
\end{align}
Using \eqref{B_varphi_exp} and $d\ell/B = \cJ d\varphi$ we see that these constraints can be written as 
\begin{align}
    \del_\psi \oint \B \cdot \bm{d\ell} + \oint \frac{d\ell}{B} F_\psi=0, \quad \del_\alpha \oint \B \cdot \bm{d\ell} + \oint  \frac{d\ell}{B} F_\alpha=0.
    \label{Newcomb}\\
     \del_\alpha  \oint \frac{d\ell}{B} F_\psi- \del_\psi  \oint \frac{d\ell}{B} F_\alpha =0 \nonumber
\end{align}
We note that the last consistency condition is not independent of the first two since it is obtained by eliminating $B_\phi$ from the first two conditions. However, once the integral constraints \eqref{Newcomb} are satisfied, we can solve the MDEs for $B_\psi$ and $B_\alpha$. Therefore, the constraints  \eqref{Newcomb} are necessary and sufficient for force-balance.

For ideal MHD equilibrium $\F=p'(\psi)\dl\psi$ and we recover \eqref{ideal_MHD_dlB_oonstraints} from \eqref{Newcomb}. For steady flows both $F_\psi$ and $F_\alpha$ are nonzero and we recover \eqref{solvability_B_circulations} and \eqref{C2prime_comstraint}.

To show the sufficiency of the integral conditions, we start with the solution of the force-balance condition and impose the divergence-free condition on the current. Writing $\J=\J_\perp + \B (\jpl/B)$, and taking the divergence, we obtain a magnetic differential equation for the parallel component of current,
\begin{align}
    \BD\lbr\frac{\jpl}{B} \rbr + \dl \cdot \J_\perp =0.
    \label{div_J_condition}
\end{align}
Therefore, if the Newcomb condition
\begin{align}
    \oint \frac{d\ell}{B} \dl\cdot \J_\perp =0
    \label{Newcomb_div_J_r}
\end{align}
is not satisfied, then current singularity follows \citep{Loizu_2015_magnetic_islands_singular_currents}. 

Using \eqref{gen_force_bal} and \eqref{Bcon_Bcov_forms} we can obtain an expression for the perpendicular component of the current
\begin{align}
\J_\perp=\frac{\B \times \F}{B^2}= \B \lbr\frac{ B_\psi F_\alpha -F_\psi B_\alpha }{B^2}\rbr +\cJ^{-1}B_\varphi \lbr F_\psi \bm{e}_\alpha -F_\alpha \bm{e}_\psi \rbr,
    \label{J_r}
\end{align}
where, $\bm{e}_\psi= \cJ \dl\alpha\times \dl \varphi$ etc are the basis vectors. The divergence of $\J_\perp$ is
\begin{align}
\dl\cdot \J_\perp = \BD \lbr \frac{1}{B^2}\lbr  B_\psi F_\alpha -F_\psi B_\alpha\rbr\rbr +\cJ^{-1}\lbr \del_\alpha (\cJ F_\psi)- \del_\psi (\cJ F_\alpha) \rbr.
    \label{div_J_r}   
\end{align}
If the integral constraints  \eqref{Newcomb} are satisfied, \eqref{MDE_the_third} is solvable and we can rewrite \eqref{div_J_r} as
\begin{align}
\dl\cdot \J_\perp = \BD \lbr \frac{1}{B^2}\lbr  B_\psi F_\alpha -F_\psi B_\alpha\rbr + \lbr \del_\alpha B_\psi-\del_\psi B_\alpha\rbr\rbr,
    \label{temp_div_J_r}   
\end{align}

which satisfies \eqref{Newcomb_div_J_r}. The MDE \eqref{div_J_condition} can then be solved for the Pfirsch-Schl\"uter currents.

Therefore, the integral constraints \eqref{Newcomb} are sufficient for the current $\J$ to satisfy force-balance and be divergence-free.

\section{Derivation of the flow-modified GGS}
\label{app:GGS}
Since the steps in deriving the flow-modified GGS are a bit tricky and one can easily get lost in the plethora of terms, we give a step-by-step derivation in this Appendix. Starting with the flow $\u$ given by \eqref{u_form} we obtain the following expression for the vorticity $\bm{\omega}$ and $\u\times \bm{\omega}$
\begin{subequations}
\begin{align}
    \bm{\omega} \equiv \dl \times \u =&\frac{\Lambda}{\rho} \J + \dl \lbr \frac{\Lambda}{\rho}\rbr \times \B -\Phi''(\psi)\dl\psi \times \Q -\Phi'(\psi)\dl \times \Q.
    \label{omega_form}\\
    \u\times \bm{\omega}=& -\frac{\Lambda}{\rho}\J\times \u-\u\cdot\dl \lbr \frac{\Lambda}{\rho}\rbr \B +\uB \dl \lbr \frac{\Lambda}{\rho}\rbr \label{uxOmega}\\
    &-\Phi''(\psi)\uQ \dl \psi -\Phi'(\psi)\u\times (\dl\times \Q).
   \nonumber
\end{align}
\label{omega_n_uxomega}
\end{subequations}
With the help of \eqref{omega_n_uxomega} we obtain the right hand side of $\dl\psi$ dotted with the vorticity equation \eqref{vorticity_eqn}
\begin{align}
\dl\psi\cdot \lbr\u\times \bm{\omega} +\frac{1}{\rho}\JxB \rbr=  \dl\psi\cdot &\left( \frac{1}{\rho}\J\times \lbr \B -\Lambda \u\rbr +\uB  \dl \lbr \frac{\Lambda}{\rho}\rbr\right. \label{temp_RHS_vort} \\
&-\bigg. \Phi''(\psi)\uQ \dl \psi \bigg) +\Phi'(\psi)\lbr \u\times\dl\psi\rbr \cdot \lbr \dl\times \Q\rbr \nonumber
\end{align}
Now, using \eqref{Bernoulli}, the left side of $\dl\psi$ dotted with the vorticity equation yields
\begin{align}
\dl\psi\cdot \lbr \frac{1}{2}u^2 +h\rbr= \dl\psi\cdot\lbr \dl\psi\lbr \frac{dH(\psi)}{d\psi} -\Phi''(\psi)\uQ\rbr+  \dl \cH -\Phi'(\psi)\dl \uQ   \rbr
    \label{temp_LHS_vort}
\end{align}
We can see that the $\Phi''(\psi)$ term cancels when we equate the left \eqref{temp_LHS_vort} to the right side \eqref{temp_RHS_vort}.

Let us first evaluate the first term on the right hand side of \eqref{temp_RHS_vort} by replacing $\dl \psi$ by $\B\times\Q$ such that
\begin{align}
\dl\psi\cdot  \J\times \lbr \B -\Lambda \u\rbr= (\J\cdot \Q) (B^2-\Lambda \uB) -(\J\cdot \Q) (\BQ-\Lambda \uQ)
    \label{JxBmLu}
\end{align}
The components of $\J$ can be calculated as follows
\begin{subequations}
\begin{align}
\J\cdot\Q&= \dl \cdot \lbr \B\times\Q \rbr+\dl \cdot \lbr \Q\times\B \rbr = \nabla^2\psi + \frac{1}{Q^2}\lbr \BQ \Q + \Q\times \dl\psi \rbr\cdot (\dl\times \Q)
\nonumber\\
&= Q^2\dl\cdot\lbr \frac{1}{Q^2}\dl\psi\rbr+\frac{1}{Q^2}\lbr \w\cdot \dl \psi + \BQ \Q\cdot \dl\times \Q\rbr \label{JQ}\\
\J\cdot\B &= \J\cdot\frac{1}{Q^2}\lbr \BQ \Q + \Q\times \dl\psi \rbr=\frac{\BQ}{Q^2}\J\cdot\Q+ \lbr \bm{F_Q}-\dl \BQ \rbr\cdot\frac{\dl\psi}{Q^2}
\label{JB}
\end{align}
\label{JB_JQ}
\end{subequations}
Substituting \eqref{JB_JQ} in \eqref{JxBmLu} and using
\begin{align}
    B^2=\frac{1}{Q^2}\lbr\BQ^2+|\dl\psi|^2\rbr, \quad \lbr \uB-\frac{\BQ\uQ}{Q^2}\rbr =\frac{\Lambda}{\rho}|\dl \psi|^2,
    \label{uBmCuQ}
\end{align}
we get
\begin{align}
    \frac{\dl\psi}{\rho}\cdot \J\times\lbr \B -\Lambda \u\rbr=&(\J\cdot\Q)\frac{|\dl\psi|^2}{\rho Q^2}\lbr 1-\frac{\Lambda^2}{\rho}\rbr+\frac{(\B-\Lambda\u)\cdot\Q}{\rho Q^2} \dl\psi\cdot \lbr\dl\BQ  +\bm{F_Q} \rbr \nonumber\\
    =& \frac{|\dl\psi|^2}{\rho}\dl \cdot \lbr\frac{1}{Q^2}\lbr 1-\frac{\Lambda^2}{\rho}\rbr\dl \psi \rbr + Y_1 \\
    &+ \frac{\dl\psi}{\rho Q^2}\cdot\lbr |\dl\psi|^2\dl \lbr \frac{\Lambda^2}{\rho}\rbr+ \lbr \BQ -\Lambda \uQ\rbr \dl\BQ \rbr \nonumber
\end{align}
where,
\begin{align}
     Y_1=\frac{1}{\rho Q^2}\lbr \frac{|\dl\psi|^2}{Q^2} \lbr 1-\frac{\Lambda^2}{\rho}\rbr \lbr \w\cdot\dl\psi + \BQ \Q\cdot \dl\times\Q \rbr +\lbr \BQ -\Lambda \uQ\rbr\dl\psi\cdot \bm{F_Q} \rbr
\end{align}

In simplifying the last term term of \eqref{temp_RHS_vort} we use instead of $\u\times \dl \psi$
\begin{align}
   \u\times (\B\times \Q)=\uQ \B -\uB \Q=& \frac{1}{Q^2}\lbr -|\dl\psi|^2 \frac{\Lambda}{Q}\Q + \uQ \Q\times \dl \psi\rbr\nonumber\\
   \therefore\quad \Phi'(\psi)\lbr \u\times \dl\psi\rbr \cdot \lbr \dl \times \Q\rbr =& -\Phi'(\psi)\frac{\dl \psi\cdot \dl Q^2}{Q^2}\uQ -Y_2\\
   Y_2=&-\Phi'(\psi)\lbr \frac{\uQ}{Q^2}\w\cdot \dl \psi - \frac{\Lambda}{\rho}\frac{\Q\cdot \dl\times \Q}{Q^2}|\dl\psi|^2\rbr \nonumber
\end{align}
We now rearrange all the terms in the $\dl\psi\cdot$ component of the vorticity equation in the form
\begin{align}
    |\dl\psi|^2\lbr \rho \frac{dH(\psi)}{d\psi}+\dl\cdot \lbr\frac{1}{Q^2}\lbr1-\frac{\Lambda^2}{\rho}\rbr\dl\psi \rbr \rbr + \rho\dl\psi\cdot \dl \cH + \rho(Y_1+Y_2 + Y_3)=0,
    \label{almost_there}
\end{align}    
where $Y_3$ is given by
\begin{align}
    Y_3= - \dl\psi\cdot\left[-\frac{1}{\rho Q^2}\lbr |\dl\psi|^2\dl \lbr \frac{\Lambda^2}{\rho} \rbr + \lbr C-\Lambda \uQ\rbr\dl C\rbr +\uB\dl \lbr\frac{\Lambda}{\rho} \rbr \right. \nonumber \\
    \left. +\Phi'(\psi)\lbr \dl \uQ -\frac{\dl Q^2}{Q^2}\uQ\rbr \right],
\end{align}
with $C=\BQ$.
We simplify $Y_3$ by using $\dl (\Lambda^2/\rho)=\Lambda\dl (\Lambda/\rho)+(\Lambda/\rho)\dl \Lambda$ and \eqref{uBmCuQ} 
\begin{align}
    \frac{1}{\rho}|\dl\psi|^2\dl\lbr \frac{\Lambda^2}{\rho}\rbr =\lbr \uB -\frac{\uQ C}{Q^2} \rbr  \lbr \dl \lbr \frac{\Lambda}{\rho}\rbr + \frac{\dl \Lambda}{\rho}  \rbr.
    \label{dl_Lambdasqr_by_rho}
\end{align}
Substituting \eqref{dl_Lambdasqr_by_rho} into $Y_3$ and simplifying we get
\begin{align}
    Y_3=\frac{|\dl\psi|^2}{\rho} \uB\frac{d\Lambda}{d\psi} +& \frac{1}{\rho} \dl\psi \cdot \left[-\frac{C}{\rho Q^2}\dl C +\frac{\uQ}{Q^2}\lbr C \dl \lbr \frac{\Lambda}{\rho}\rbr+\frac{C}{\rho}\dl \Lambda + \frac{\Lambda}{\rho}\dl C \rbr \right.\nonumber\\
    \bigg. &+\Phi'(\psi)\lbr \dl \uQ -\frac{\dl Q^2}{Q^2}\uQ\rbr\bigg]
    \label{Y_3simp}
\end{align}
The final step is to use the generalized momentum equation \eqref{Q_momentum} to replace the term, $C\dl C$ in \eqref{Y_3simp} by $C\dl \lbr I(\psi)+\cI +\Lambda \uQ\rbr$. After some straightforward algebra we get
\begin{align}
    \rho Y_3 = |\dl\psi|^2 \lbr \uB \frac{d\Lambda}{d\psi}+ \frac{\BQ}{Q^2}\frac{dI}{d\psi}-\rho \frac{d\Phi'(\psi)}{d\psi}\uQ\rbr +\BQ \dl\psi\cdot \dl\cI  
    \label{Y3}
\end{align}
Substituting \eqref{Y3} into \eqref{almost_there} we get \eqref{GGS}.

\section{Derivation of the final Hamada condition}\label{app:derivation_final_Hamada}

To derive the final Hamada condition \eqref{C2prime_comstraint}, it is convenient to set up flux-coordinates $(\psi,\alpha,\beta)$ (discussed in \citep{hameiri1998variational}) such that
\begin{align}
    \B=\dl\psi\times \dl \alpha , \quad \Q = \dl\psi\times \dl \beta.
    \label{Hameri_coordinates}
\end{align}
For closed field line systems with zero magnetic shear, $\dl\alpha$ is a single-valued quantity. Since, $\B$ and $\Q$ must be single-valued, $\dl\beta$ in general is either single-valued or at least the multi-valued part of $\dl\beta$ is in the $\dl\psi$ direction \citep{hameiri1998variational}. For closed $\Q$ lines, $\beta$ is single-valued. 

Although we use the $(\psi,\alpha,\beta)$ coordinates to derive \eqref{C2prime_comstraint}, the result is independent of the choice and details of the flux coordinates. In particular, any flux-coordinates that allow radial currents, e.g., generalized Boozer coordinates  \citep{Rodriguez2020a} can be used.

The condition $\B\times\Q=\dl \psi$ implies that the Jacobian $\cJ^{-1}=\dl\psi\times \dl\alpha \cdot \dl \beta$ is unity. Therefore,
\begin{align}
    \BD = \del_\beta, \quad \QD = -\del_\alpha.
\end{align}
In these coordinates, the divergence of a vector of the form $f\dl \psi$ is given by
\begin{align}
    \dl\cdot \lbr f \dl\psi \rbr = \del_\psi \lbr f |\dl\psi|^2\rbr - \QD \lbr f \dl\psi\cdot \dl \alpha\rbr +\BD \lbr f \dl\psi\cdot \dl \beta\rbr.
    \label{proxy_div}
\end{align}
Averaging \eqref{proxy_div} along the closed magnetic field line we get
\begin{align}
    \langle  \dl\cdot \lbr f \dl\psi \rbr\rangle = \del_\psi  \langle  f |\dl\psi|^2 \rangle  - \QD \langle f \dl\psi\cdot \dl \alpha\rangle +f |\dl\psi|^2[\beta],
    \label{avg_proxy_div}
\end{align}
where,  $[\beta]$ denotes the jump in $\beta$ along the closed field line. If $\Q$ lines are closed $[\beta]=0$. 

We obtain \eqref{C2prime_comstraint} for $f=(1-\Lambda^2/\rho)/Q^2$ upon using \eqref{Bernouili_n_Q_momentum},\eqref{solvability_u_B_circulations} and the identity
\begin{align}
    \lbr 1-\frac{\Lambda^2}{\rho}\rbr \frac{1 }{Q^2} |\dl \psi|^2 = B^2 -\Lambda \uB -\frac{\BQ}{Q^2}(I+\cI).
\end{align}

 \bibliographystyle{jpp}
 \bibliography{plasmalit}

\end{document}